\documentclass[twocolumn,showpacs,preprintnumbers,amsmath,amssymb]{revtex4}
\usepackage[dvips]{graphicx}% Include figure files
\usepackage{dcolumn}% Align table columns on decimal point
\usepackage{bm}% bold math

\newcommand{\ovec}[1]{{\mbox{\boldmath $#1$}}}

\newcommand{\bA}{\ovec{A}}
\newcommand{\bb}{\ovec{b}}
\newcommand{\bB}{\ovec{B}}

\newcommand{\bD}{\ovec{D}}

\newcommand{\bscE}{\ovec{\cal{E}}}

\newcommand{\bff}{\ovec{f}}
\newcommand{\bg}{\ovec{g}}

\newcommand{\bJ}{\ovec{J}}
\newcommand{\bk}{\ovec{k}}
\newcommand{\bK}{\ovec{K}}

\newcommand{\br}{\ovec{r}}
\newcommand{\bR}{\ovec{R}}
\newcommand{\bu}{\ovec{u}}
\newcommand{\bU}{\ovec{U}}
\newcommand{\bv}{\ovec{v}}
\newcommand{\bw}{\ovec{w}}
\newcommand{\bW}{\ovec{W}}
\newcommand{\bx}{\ovec{x}}

\newcommand{\balpha}{\ovec{\alpha}}
\newcommand{\bbeta}{\ovec{\beta}}
\newcommand{\bgamma}{\ovec{\gamma}}
\newcommand{\bdelta}{\ovec{\delta}}
\newcommand{\bkappa}{\ovec{\kappa}}

\newcommand{\bOmega}{\ovec{\Omega}}
\newcommand{\mB}{\overline{B}}

\newcommand{\bmB}{\overline{\ovec{B}}}

\newcommand{\bmU}{\overline{\ovec{U}}}
\newcommand{\bnab}{{\mbox{\boldmath $\nabla$}}}
\def\p {\partial}
\def\dd {\mbox{d}}
\def\hb {\hat{b}}
\def\hu {\hat{u}}
\def\bzo {{\bf 0}}
\def\iu {\mbox{i}}
\def\x {\times}
\def\ol {\overline}
\def\te {\tilde{\eta}}
\def\tn {\tilde{\nu}}
\def\o {\omega}

\begin{document}

\preprint{APS/123-QED}

\title{The mean electromotive force due to turbulence of a conducting fluid\\
in the presence of mean flow}

\author{Karl-Heinz R\"adler}
 \email{khraedler@arcor.de}
\affiliation{Astrophysical Institute Potsdam, An der Sternwarte 16, D-14482 Potsdam, Germany}
\author{Rodion Stepanov}
 \email{rodion@icmm.ru}
\affiliation{Institute of Continuous Media Mechanics, Korolyov street 1, 614061 Perm, Russia}

\date{\today}% It is always \today, today,

\begin{abstract}
The mean electromotive force caused by turbulence of an
electrically conducting fluid, which plays a central part in
mean--field electrodynamics, is calculated for a rotating fluid.
Going beyond most of the investigations on this topic, an
additional mean motion in the rotating frame is taken into account.
One motivation for our investigation originates from
a planned laboratory experiment with a Ponomarenko-like dynamo.
In view of this application the second--order correlation approximation is used.
The investigation is of high interest in astrophysical context, too.
Some contributions to the mean electromotive are revealed
which have not been considered so far,
in particular contributions to the $\alpha$--effect and related effects
due to the gradient of the mean velocity.
Their relevance for dynamo processes is discussed.
In a forthcoming paper the results reported here will be specified
to the situation in the laboratory
and partially compared with experimental findings.
\end{abstract}

\pacs{47.65.-d, 47.27.-i}% PACS, the Physics and Astronomy
\maketitle

\section{Introduction}

In mean--field electrodynamics of turbulent fluids the mean
electromagnetic fields follow Maxwell's equations.
The turbulence, however, gives rise to a mean electromotive force,
which occurs in Ohm's law and, consequently, in the induction equation.
This mean electromotive force, which is crucial in the theory of cosmic magnetic fields
and dynamos as well as in other fields, has been an objective of many investigations.
It has been calculated in specific approximations for different forms of turbulence
on a rotating body under the assumption of zero mean motion of the
fluid in the rotating frame,
see, e.g., \cite{krauseetal71b,krauseetal80,raedler76,raedler80,raedler00c,kichatinov82,vainshteinetal83,
ruedigeretal93c,kitchatinovetal94b,raedleretal03}.
In a few cases also the effect of a mean motion has been studied.
There are some rather general results of that kind,
e.g., \cite{krauseetal71b,krauseetal80},
the application of which requires however further elaboration.
The more detailed results derived recently,
\cite{urpin99,urpin99c,hoyng03} and \cite{rogachevskiietal03},
are not in convincing agreement with each other.

By this reason we have again dealt with the mean electromotive force in a
rotating turbulent fluid in the presence of a mean motion.
The primary motivation for dealing with this topic was to find estimates of the effects of turbulence
in a laboratory experiment with a screw dynamo as proposed by Ponomarenko \cite{ponomarenko73},
which is under preparation in the Institute for Continuous Media Mechanics in Perm;
see \cite{denisovetal99,fricketal02,fricketal02b,noskovetal04}.
Moreover the results are of high interest for astrophysical applications,
for instance in view of the possibility of the ``$\bW \x \bJ$ dynamo",
which has been proposed recently
\cite{rogachevskiietal03,rogachevskiietal04}.

In this paper the mean electromotive force is considered in the
presence of a more or less arbitrary mean flow, and in a
forthcoming one \cite{raedleretal06b} we will specify the results and discuss
them in view of the situation in the experimental device.
(For a first, very short report on this topic see \cite{raedleretal04}.)
In Section 2 of this paper we describe the general framework of our
investigation. In Section 3 we explain some general aspects of our
view on the problem and use basic symmetry laws to draw
conclusions concerning the structure of the mean electromotive
force, that is, concerning its dependence on vectorial and
tensorial quantities that characterize the turbulence and the mean
motion. In order to determine the mean electromotive force
completely, we introduce in Section 4 specific approximations, in
particular some kind of second--order approximation, and calculate
all of its coefficients in their dependence on the intensity of
the turbulence and related parameters. Finally in Section 5 we
discuss our results in general terms, compare them with those
of other investigations and point out their consequences
for dynamo processes.

\section{Mean--field magnetohydrodynamics}
\label{sec:mfedyn}

\subsection{Electromagnetic fundamentals}
\label{subsec:mf1}

Let us assume that the behavior of the magnetic field $\bB$ in an electrically conducting fluid
is governed by the induction equation
\begin{equation}
\partial_t \bB - \bnab \x (\bU \x \bB) - \eta \bnab^2 \bB = {\bf 0}\, , \quad
   \bnab \cdot \bB = 0  \, .
\label{eq01}
\end{equation}
$\bU$ is the velocity and $\eta$ the magnetic diffusivity of the fluid,
the latter being considered as constant.

Following the lines of mean-field electrodynamics
(see, e.g., \cite{krauseetal80,raedler00c})
we define mean magnetic and velocity fields, $\bmB$ and $\bmU$,
as averages over space or time scales larger than those of the turbulence.
We call $\bB - \bmB$ and $\bU - \bmU$ simply ``fluctuations"
and denote them by $\bb$ and $\bu$, respectively.
We further assume that the Reynolds averaging rules apply.
Taking the average of equation (\ref{eq01}) we obtain the
mean-field induction equation
\begin{equation}
\partial_t \bmB - \bnab \x (\bmU \x \bmB + \bscE)
   - \eta \bnab^2 \bmB = {\bf 0} \, , \quad
   \bnab \cdot \bmB = 0  \, ,
\label{eq03}
\end{equation}
where $\bscE$ is the mean electromotive force due to the fluctuations of magnetic field
and velocity,
\begin{equation}
\bscE = \ol{\bu \x \bb} \, .
\label{eq05}
\end{equation}

The equation for $\bb$ resulting from (\ref{eq01}) and (\ref{eq03})
allows us to conclude that $\bscE$ can be considered as a functional
of $\bmU$, $\bu$ and $\bmB$, which is linear in $\bmB$.
Furthermore $\bscE$ in a given point in space and time depends on $\bmU$, $\bu$ and $\bmB$
not only in this point but also on their behaviors in a certain neighborhood of this point.
We assume that $\bscE$ has no part independent of $\bmB$, that is, it is not only linear
but also homogeneous in $\bmB$.
We further accept the frequently used assumption that $\bmB$ varies only weakly in space and time
so that $\bscE$ in a given point depends on $\bmB$ only via its
components and their first spatial derivatives in this point.
Hence, $\bscE$ can be represented in the form
\begin{equation}
{\cal{E}}_i = a_{ij} \, \overline{B}_j
 + b_{ijk} \, \partial \overline{B}_j / \partial x_k \, ,
\label{eq07}
\end{equation}
where the tensors $a_{ij}$ and $b_{ijk}$ are averaged quantities determined
by $\bmU$ and $\bu$.
Here and in the following a Cartesian co--ordinate system $(x_1, x_2, x_3)$
is used and the summation convention is adopted.
Relation (\ref{eq07}) is equivalent to
\begin{eqnarray}
&& \!\!\!\!\!\!
   \bscE = - \balpha \circ \bmB - \bgamma \x \bmB
\nonumber\\
&& \quad
   - \bbeta \circ (\bnab \x \bmB) - \bdelta \x (\bnab \x \bmB)
   - \bkappa \circ (\bnab \bmB)^{(s)} \, ; \quad
\label{eq09}
\end{eqnarray}
see, e.g., \cite{raedler80} or \cite{raedleretal03}.
Here $\balpha$ and $\bbeta$ are symmetric tensors of the second rank,
$\bgamma$ and $\bdelta$ are vectors,
and $\bkappa$ is a tensor of the third rank,
all depending on $\bmU$ and $\bu$ only.
Further $(\bnab \bmB)^{(s)}$ is the symmetric part of the gradient tensor of $\bmB$,
i.e. $(\bnab \bmB)^{(s)}_{ij} = \frac{1}{2}
(\partial \ol{B}_i / \partial x_j + \partial \ol{B}_j / \partial x_i)$.
Notations like $\balpha \circ \bmB$ are used
in the sense of $(\balpha \circ \bmB)_i = \alpha_{ij} {\ol{B}}_j$,
and $\bkappa \circ (\bnab \bmB)^{(s)}$ is defined by
$(\bkappa \circ (\bnab \bmB)^{(s)})_i = \kappa_{ijk} (\bnab \bmB)^{(s)}_{jk}$.

The term with $\balpha$ in (\ref{eq09}) describes the $\alpha$--effect,
which is in general anisotropic,
that with $\bgamma$ a transport of mean magnetic flux by the turbulence.
The terms with $\bbeta$ and $\bdelta$ can be interpreted by introducing
a modified magnetic diffusivity, again in general anisotropic.
The induction effects which correspond to these terms are usually accompanied
by such described by the term $\bkappa$, which allows no simple independent interpretation.
More details will be explained in Sections~\ref{subsecform}--\ref{subsecdelta}.

The quantities  {$\balpha$, $\bgamma$, $\bbeta$, $\bdelta$ and
$\bkappa$} are connected with $a_{ij}$ and $b_{ijk}$ by
\begin{eqnarray}
&& \alpha_{ij} = - \frac{1}{2} (a_{ij} + a_{ji}) \, , \quad
   \gamma_i = \frac{1}{2} \epsilon_{ijk} a_{jk}
\nonumber\\
&& \beta_{ij} = \frac{1}{4} (\epsilon_{ikl} b_{jkl} + \epsilon_{jkl} b_{ikl}) \, , \quad
   \delta_i = \frac{1}{4} (b_{jji} - b_{jij}) \, , \quad
\label{eq11}\\
&& \kappa_{ijk} = - \frac{1}{2} (b_{ijk} + b_{ikj}) \, .
\nonumber
\end{eqnarray}

\subsection{Momentum balance}

We will consider the situation as described so far in a rotating frame
of reference and restrict our attention to an incompressible fluid.
The fluid velocity $\bU$ is assumed to satisfy the momentum balance
and the incompressibility condition in the form
\begin{eqnarray}
\partial_t \bU + (\bU \cdot \bnab) \bU &=& \varrho^{-1} \bnab p + \nu \bnab^2 \bU
   - 2 \bOmega \x \bU + \bff
\nonumber\\
&& \qquad \qquad \bnab \cdot \bU = 0 \, .
\label{eq18}
\end{eqnarray}
Here $\varrho$ is the mass density and $\nu$ the kinematic viscosity of the fluid,
$p$ the hydrodynamic pressure including the centrifugal pressure,
$\bOmega$ the angular velocity responsible for the Coriolis force,
and $\bff$ an artificial external force,
which should mimic the cause of the turbulence.
Any influence of the magnetic field on the fluid motion is ignored.

\section{The structure of the mean electromotive force $\bscE$
%%in the presence of a mean motion
}

\subsection{Change to a proper frame of reference}

Let us now focus our attention on the electromotive force $\bscE$ in
a given point, consider the mean motion as independent of time and
specify the frame of reference in which (\ref{eq18}) applies such
that $\bmU = {\bf 0}$ in this point. { $\bscE$ has to be interpreted
as a force on charged particles rather than a part of the electric
field. Therefore the result for $\bscE$ obtained in a given frame},
understood as a vector with the usual transformation properties,
applies then also in any other frame; see also
\cite{raedleretal06b}. Remaining in the frame specified such that
$\bU = \bzo$ in the considered point we introduce the simplifying
assumption that in the neighborhood of this point relevant for the
determination of $\bscE$ the mean velocity $\bmU$ varies only
weakly. More precisely, we assume that it can be represented there
with respect to the frame specified above in the form ${\ol{U}}_i =
U_{ij} x_j$ with $U_{ij}$ being constant, where $(x_1, x_2, x_3)$
means a new Cartesian co--ordinate system defined in the rotating
frame such that $x_1 = x_2 = x_3 = 0$ in the point considered.

\subsection{Homogeneous background turbulence}

We further assume until further notice that the turbulent fluctuations $\bu$
deviate from a homogeneous isotropic mirror--symmetric turbulence
only as a consequence of the Coriolis force defined by $\bOmega$
and of the gradient of the mean fluid velocity,
that is, the gradient tensor $\bnab \bmU$
given by $(\bnab \bmU)_{ij} = \partial {\ol{U}}_i / \partial x_j$,
or $(\bnab \bmU)_{ij} = U_{ij}$.
For particular purposes we split $\bnab \bmU$ into its symmetric
and antisymmetric parts.
The symmetric one is the rate of strain tensor, $\bD$,
given by $D_{ij} = \frac{1}{2}
(\partial {\ol{U}}_i / \partial x_j + \partial {\ol{U}}_j / \partial x_i)$.
It describes the deforming motion close to the point considered.
Due to the incompressibility of the fluid we have $\bnab \cdot \bmU = 0$
and therefore $D_{ii} = 0$.
The antisymmetric part, $\bA$,
given by $A_{ij} = \frac{1}{2}
(\partial {\ol{U}}_i / \partial x_j - \partial {\ol{U}}_j / \partial x_i)$,
corresponds to a rigid body rotation of the fluid close to this point.
We may represent it according to $A_{ij} = - \frac{1}{2}\epsilon_{ijl} W_l$
by the vector $\bW = \bnab \x \bmU$.

In order to prepare conclusions concerning the structure of $\bscE =
\overline{\bu \x \bb}$ we first recall symmetry properties of
the equations (\ref{eq01}) and (\ref{eq18}) governing $\bB$ and $\bU$.
If these equations are satisfied with given $\bB$, $\bU$, $\bnab p$,
$\bOmega$ and $\bff$, they are, too, with other $\bB$, $\bU$, $\bnab
p$, $\bOmega$ and $\bff$ derived from the given ones by a
rotation about any axis running, e.g., through $\bx = {\bf 0}$.
Likewise they are satisfied with $\bB$, $\bU$, $\bnab p$,
$\bOmega$ and $\bff$ derived from the given ones by reflecting them at
a plane, e.g., one containing $\bx = {\bf 0}$ and, in addition,
changing the signs of $\bB$ and $\bOmega$.
These two properties apply analogously to consequences drawn from these equations,
in particular to the equations governing $\bu$ and $\bb$.
The first property, connected with the rotation of fields, leads to
the conclusion that the tensors $a_{ij}$ and $b_{ijk}$, which occur
in (\ref{eq07}), and therefore also $\balpha$, $\bgamma$,
$\bbeta$, $\bdelta$ and $\bkappa$ cannot contain any
construction elements other than the isotropic tensors
$\delta_{lm}$ and $\epsilon_{lmn}$, the vectors $\bOmega$ and
$\bW$ and the tensor $\bD$.
Note that the force $\bff$, which is assumed to cause
a homogeneous isotropic turbulence, can not introduce other than
isotropic quantities.
The second property, connected with reflection, is often described
by considering $\bU$, $\bnab p$, and $\bff$ as polar and
$\bB$ and $\bOmega$ as axial vector fields.
By contrast to polar vectors the axial ones show specific sign changes
of their components under reflection of the coordinate system.
We adopt this concept and distinguish between ``true" and ``pseudo"
scalars, vectors and tensors, where we call polar and axial
vectors simply ``true" and ``pseudo" vectors, respectively.
Then $a_{ij}$ and $b_{ijk}$ have to be pseudo tensors.
This implies that $\balpha$, $\bdelta$ and $\bkappa$
are also pseudo quantities but $\bgamma$ and $\bbeta$
true quantities.

Let us consider first $\balpha$ and $\bgamma$.
The mentioned construction elements $\delta_{lm}$, $\epsilon_{lmn}$, $\bOmega$, $\bW$ and $\bD$
allow us neither to built a pseudo tensor of the second rank nor a true vector.
That is, we have
\begin{equation}
\alpha_{ij} = 0 \, , \quad \gamma_i = 0 \, .
\label{eq19}
\end{equation}

In contrast to this there are several non--zero contributions to $\beta_{ij}$,
$\delta_i$ and $\kappa_{ijk}$.
For the sake of simplicity we give only those of them which are linear
in $\bOmega$, $\bW$ and $\bD$, that is,
\begin{eqnarray}
&& \beta_{ij} = \beta^{(0)} \delta_{ij} + \beta^{(D)} D_{ij} \, , \quad
    \delta_i = \delta^{(\Omega)} \Omega_i + \delta^{(W)} W_i  \, , \quad
\nonumber\\
&& \kappa_{ijk} = {\frac{1}{2}} \kappa^{(\Omega)} (\Omega_j
\delta_{ik} + \Omega_k \delta_{ij})
\label{eq21}\\
&& \quad
    + {\frac{1}{2} }\kappa^{(W)} (W_j \delta_{ik} + W_k \delta_{ij})
    + \kappa^{(D)} (\epsilon_{ijl} D_{kl} + \epsilon_{ikl} D_{jl}) \, .
\nonumber
\end{eqnarray}
Here $\beta^{(0)}$, $\beta^{(D)}$, $\delta^{(\Omega)}$, $\cdots$ are
coefficients determined by $\bu$ but independent of $\bOmega$, $\bW$
and $\bD$. Because of $\bnab \cdot \bmB = 0$ terms of $\kappa_{ijk}$
containing $\delta_{jk}$ would not contribute to $\bscE$ and may
therefore be dropped.

As a consequence of (\ref{eq19}) and (\ref{eq21}) we have
\begin{eqnarray}
\bscE &=& - \beta^{(0)} \bnab \x \bmB
    - \beta^{(D)} \bD \circ (\bnab \x \bmB)
\nonumber\\
&& - (\delta^{(\Omega)} \bOmega + \delta^{(W)} \bW) \x (\bnab \x \bmB)
\label{eq23}\\
&& \!\!\!\!\!\!\!\!\!\!\!\!\!\!\!
- (\kappa^{(\Omega)} \bOmega + \kappa^{(W)} \bW) \circ (\bnab \bmB)^{(s)}
%%\nonumber\\
%%&& \qquad
   - \kappa^{(D)} \, \hat{\bkappa} (\bD) \circ (\bnab \bmB)^{(s)} \, ,
\nonumber
\end{eqnarray}
where $\hat{\bkappa} (\bD)$ is a tensor of the third rank
defined by ${\hat{\kappa}}_{ijk} = \epsilon_{ijl} D_{lk} + \epsilon_{ikl} D_{lj}$.
Quantities like $\beta^{(0)}$, $\beta^{(D)}$, $\cdots$ $\kappa^{(D)}$ are called
``mean--field coefficients" in the following.

The $\beta^{(0)}$ and $\beta^{(D)}$ terms in (\ref{eq23}) make
that the mean--field diffusivity deviates from the original magnetic diffusivity $\eta$
of the fluid.
Due to the $\beta^{(0)}$ term the mean--field diffusivity turns into $\eta + \beta^{(0)}$,
due to the $\beta^{(D)}$ term it becomes anisotropic.
The $\delta^{(\Omega)}$ and $\delta^{(W)}$ terms, too, can be discussed as contributions
to the mean--field diffusivity.
They lead to skew--symmetric contributions to the diffusivity tensor.
In another context the effect described by the $\delta^{(\Omega)}$ term
has been called ``$\bOmega \x \bJ$--effect".
It has been shown that this effect in combination with a differential rotation,
here a dependence of $\bOmega$ on $r$,
is able to establish a dynamo; see \cite{raedler69b,raedler86,roberts72,moffattetal82}.
The $\delta^{(W)}$ term describes an effect analogous to the $\bOmega \x \bJ$--effect,
which has been revealed only recently \cite{rogachevskiietal03}.
We call it ``$\bW \x \bJ$--effect".
It occurs however even in the absence of the Coriolis force,
only as consequence of a shear in the mean motion.
We will discuss the $\delta^{(\Omega)}$ and $\delta^{(W)}$--effects
as well as the $\kappa^{(\Omega)}$, $\kappa^{(W)}$ and $\kappa^{(D)}$--effects
in more detail in Section~\ref{subsecdelta}.

\subsection{Inhomogeneous background turbulence}

Let us now relax the assumption that the original turbulence is
homogeneous and isotropic.
We admit an inhomogeneity and an anisotropy due to a gradient
of a quantity like the turbulence intensity and introduce a vector $\bg$
in the direction of this gradient,
e.g., by putting $\bnab \ol{u^2} = \bg \, \ol{u^2}$,
with $\ol{u^2}$ derived from the turbulent velocity $\bu$.
Then we have to add $\bg$ to the above--mentioned construction elements
of $\balpha$, $\bgamma$, $\bbeta$, $\bdelta$ and $\bkappa$.
As a consequence $\balpha$ and $\bgamma$ can well be non-zero.
For the sake of simplicity we assume that the influence of $\bg$
on these quantities is so weak that they are at most of first order in $\bg$.
We have then
\begin{eqnarray}
\alpha_{ij} &=& \alpha_1^{(\Omega)} (\bg \cdot \Omega) \delta_{ij}
    + \alpha_2^{(\Omega)} (g_i \Omega_j + g_j \Omega_i)
\nonumber\\
&& \quad + \alpha_1^{(W)} (\bg \cdot \bW) \delta_{ij}
    + \alpha_2^{(W)} (g_i W_j + g_j W_i)
\nonumber\\
&& \quad  + \alpha^{(D)} (\epsilon_{ilm} D_{jl} + \epsilon_{jlm} D_{il}) \, g_m
\label{eq25}\\
\gamma_i &=& \gamma^{(0)} g_i
    + \gamma^{(\Omega)} \epsilon_{ilm} g_l \Omega_m
    + \gamma^{(W)} \epsilon_{ilm} g_l W_m
\nonumber\\
&& \quad  + \gamma^{(D)} g_j D_{ij} \, ,
\nonumber
\end{eqnarray}
whereas (\ref{eq21}) remains unchanged.

Consequently $\bscE$ takes the form
\begin{eqnarray}
\bscE \!\!&=& \!\! - \alpha_1^{(\Omega)} (\bg \cdot \bOmega) \bmB
    - \alpha_2^{(\Omega)} ((\bOmega \cdot \bmB) \bg + (\bg \cdot \bmB) \bOmega)
\nonumber\\
&&\!\!  - \alpha_1^{(W)} (\bg \cdot \bW) \bmB
    - \alpha_2^{(W)} ((\bW \cdot \bmB) \bg + (\bg \cdot \bmB) \bW)
\nonumber\\
&&\!\!  - \alpha^{(D)} \hat{\balpha}(\bg, \bD) \circ \bmB
\nonumber\\
&&\!\!  - (\gamma^{(0)} \bg
\label{eq27}\\
&&\!\! \qquad
    + \gamma^{(\Omega)} \bg \x \bOmega
    + \gamma^{(W)} \bg \x \bW
    + \gamma^{(D)} \bg \circ \bD) \x \bmB
\nonumber\\
&&\!\!  - \beta^{(0)} \bnab \x \bmB
    - \beta^{(D)} \bD \circ (\bnab \x \bmB)
\nonumber\\
&&\!\!  - (\delta^{(\Omega)} \bOmega + \delta^{(W)} \bW) \x (\bnab \x \bmB)
\nonumber\\
&&\!\!\!\!\!\!\!\!\!\!\!\!\!\!\!
    - (\kappa^{(\Omega)} \bOmega + \kappa^{(W)} \bW) \circ (\bnab \bmB)^{(s)}
- \kappa^{(D)} \, \hat{\bkappa} (\bD) \circ (\bnab \bmB)^{(s)} \, ,
\nonumber
\end{eqnarray}
where $\hat{\balpha}(\bg, \bD)$ is a symmetric tensor defined by
${\hat{\alpha}}_{ij} = (\epsilon_{ilm} D_{lj} + \epsilon_{jlm} D_{li}) g_m$.

\section{Calculation of the mean electromotive force $\bscE$}
\label{sec4}

\subsection{Basic equations and approximations}
\label{sec41}

Our considerations on the structure of the electromotive force $\bscE$
did not provide us with results for the coefficients
$\alpha_1^{(\Omega)}$, $\alpha_2^{(\Omega)}$, $\cdots$, $\kappa^{(D)}$
showing their dependence on the intensity or other parameters of the turbulent flow.
To calculate these coefficients we start again from the induction
equation (\ref{eq01}) and the momentum balance in the form (\ref{eq18}),
both related to the moving frame of reference defined above.
From equation (\ref{eq01}) and its mean--field version (\ref{eq03})
and from equation (\ref{eq18}) and the corresponding mean--field version
we derive the equations governing the magnetic and velocity
fluctuations $\bb$ and $\bu$,
\begin{eqnarray}
\partial_t \bb - \eta \bnab^2 \bb
    &=&   \bnab \x (\bmU \x \bb + \bu \x \bmB + (\bu \x \bb)')
\nonumber\\
\quad \quad \bnab \cdot \bb &=& 0
\nonumber\\
\partial_t \bu - \nu \bnab^2 \bu
    &=& - \varrho^{-1} \bnab p' - 2 \bOmega \x \bu
\label{eq103}\\
&& \!\!\!\!\!\!\!\!\!\!\!\!\!\!\!\!\!\!\!\!\!\!\!\!
     - (\bmU \cdot \bnab) \bu - (\bu \cdot \bnab) \bmU - ((\bu \cdot \bnab) \bu)' + \bff'
\nonumber\\
\bnab \cdot \bu &=& 0 \, ,
\nonumber
\end{eqnarray}
where $(\bu \x \bb)' = \bu \x \bb - \overline{\bu \x \bb}$
and $((\bu \cdot \bnab) \bu)' = (\bu \cdot \bnab) \bu
- \overline{(\bu \cdot \bnab) \bu}$.
In view of the calculation of the electromotive force $\bscE$
in the point $\bx = {\bf 0}$ of the co--moving frame of reference
we consider these equations in some surroundings of this point.
Adopting the assumptions introduced above on sufficiently weak variations
of $\bmB$ and $\bmU$ in space and time we put
\begin{equation}
{\overline{B}}_i = B_i + B_{ij} \, x_j \, , \quad
     {\overline{U}}_i = U_{ij} \, x_j \, ,
\label{eq105}
\end{equation}
with $B_i$, $B_{ij}$ and  $U_{ij}$ being constants and satisfying
$U_{ii} = B_{ii} = 0$.

We restrict our calculation on the case in which the influences
of both the Coriolis force and the shear of the mean motion on $\bu$ and $\bb$
are only weak.
We introduce the expansions
\begin{equation}
\bu = \bu^{(0)} + \bu^{(1)} + \cdots \, , \quad
     \bb = \bb^{(0)} + \bb^{(1)} + \cdots \, ,
\label{eq107}
\end{equation}
where $\bu^{(0)}$ and $\bb^{(0)}$ {are independent of $\bOmega$,
$\bW$ and $\bD$, further $\bu^{(1)}$ and $\bb^{(1)}$ are of first
order and all contributions indicated by $\cdots$ are} of higher
order in these quantities. In that sense we have
\begin{eqnarray}
\bscE &=& \bscE^{(0)} + \bscE^{(1)} + \cdots
\nonumber\\
\bscE^{(0)} &=& \bscE^{(00)} \, , \quad
    \bscE^{(1)} = \bscE^{(10)} + \bscE^{(01)} \, , \,\, \cdots
\label{eq109}\\
&& \qquad \qquad  \bscE^{(\alpha \beta)} = \ol{\bu^{(\alpha)} \x \bb^{(\beta)}} \, .
\nonumber
\end{eqnarray}
In the following we restrict our attention on the case in which $\bscE$
is linear in $\bOmega$, $\bW$ and $\bD$, that is, on the terms
$\bscE^{(0)}$ and $\bscE^{(1)}$ in this expansion of $\bscE$.

We assume that both $\bu$ and $\bb$ are small enough
so that the second--order correlation approximation (SOCA) applies,
sometimes also labelled as first--order smoothing approximation (FOSA),
which is often used in astrophysical context.
So we conclude from (\ref{eq103}) that
\begin{eqnarray}
\partial_t \bu^{(0)} - \nu \bnab^2 \bu^{(0)}
      &=& - \varrho^{-1} \bnab p^{(0)}
\nonumber\\
&& \!\!\!\!\!\!\!\!\!\!\!\!\!\!\!\!\!\!\!\!\!\!\!\!
     + ((\bu^{(0)} \cdot \bnab) \bu^{(0)})' + \bff^{(0)}
\nonumber\\
\quad \quad \bnab \cdot \bu^{(0)} &=& 0
\nonumber\\
\partial_t \bu^{(1)} - \nu \bnab^2 \bu^{(1)}
      &=& - \varrho^{-1} \bnab p^{(1)}
\nonumber\\
&& \!\!\!\!\!\!\!\!\!\!\!\!\!\!\!\!\!\!\!\!\!\!\!\!
      - (\bmU \cdot \bnab) \bu^{(0)} - (\bu^{(0)} \cdot \bnab) \bmU
      - 2 \bOmega \x \bu^{(0)}
\nonumber\\
\quad \quad \bnab \cdot \bu^{(1)} &=& 0
\label{eq111}\\
\partial_t \bb^{(0)} - \eta \bnab^2 \bb^{(0)}
       &=& \bnab \x (\bu^{(0)} \x \bmB)
\nonumber\\
\quad \quad \bnab \cdot {\bb}^{(0)} &=& 0
\nonumber\\
\partial_t \bb^{(1)} - \eta \bnab^2 \bb^{(1)}
       &=& \bnab \x (\bu^{(1)} \x \bmB + \bmU \x \bb^{(0)})
\nonumber\\
\quad \quad \bnab \cdot {\bb}^{(1)} &=& 0 \, . \nonumber
\end{eqnarray}
We consider the turbulent fluid motion in the limit of zero Coriolis
force and zero shear, that is $\bu^{(0)}$, as given. In deriving
(\ref{eq111}) we have assumed that the force $\bff'$ does not depend
on $\bOmega$, $\bW$ or $\bD$ and therefore possesses no other
contributions than $\bff^{(0)}$. Following the traditional
second--order approximation as used in situations in which $\bu$ is
given we have neglected $(\bu^{(0)} \x \bb^{(0)})'$ in comparison
with $\bu^{(0)} \x \bmB$. In the same spirit we have further
neglected $((\bu^{(0)} \cdot \bnab) \bu^{(1)})' +  ((\bu^{(1)} \cdot
\bnab) \bu^{(0)})'$ in comparison with $(\bmU \cdot \bnab) \bu^{(0)}
+  (\bu^{(0)} \cdot \bnab) \bmU - 2 \bOmega \x \bu^{(0)}$ and
$(\bu^{(0)} \x \bb^{(1)})' + (\bu^{(1)} \x \bb^{(0)})'$ in
comparison with $\bu^{(1)} \x \bmB + \bu^{(1)} \x \bb^{(0)}$. {The
justification for these omissions has to be checked in all
applications.}

In Section~\ref{subsec:mf1} we have introduced the assumption that $\bscE$
has no contribution independent of $\bmB$.
In the second--order correlation approximation this assumption is automatically satisfied,
a contribution of this kind can not occur.
The second--order correlation approximation in the above sense also excludes any kind
of magnetohydrodynamic turbulence.
In the limit of small $\bmB$ the turbulence is purely hydrodynamic.

\subsection{Fourier representation}

We will carry out some of our calculations in the Fourier space.
The Fourier transformation is defined in the form
\begin{equation}
Q (\bx, t) = \int\!\!\!\int \hat{Q} (\bk, \omega)
   \exp(\iu(\bk \cdot \bx - \omega t)) \, \dd^3 k \, \dd \omega \, ,
\label{eq121}
\end{equation}
where the integrations are over all $\bk$ and $\omega$.

Let us consider the two--point correlation tensor $\phi_{ij}$
for two vector fields $\bv$ and $\bw$ defined by
\begin{equation}
\phi_{ij} (\bx_1, t_1; \bx_2, t_2)
    = \langle v_i (\bx_1, t_1) \, w_j (\bx_2, t_2) \rangle \, .
\label{eq125}
\end{equation}
Here and in what follows the notation $\langle X \rangle$ is used
in the same sense as $\ol{X}$.
Following \cite{robertsetal75} we consider $\phi_{ij}$ also as a function of the variables
\begin{eqnarray}
\bR &=& (\bx_1 + \bx_2)/2  \, , \quad \br = \bx_1 - \bx_2
\nonumber\\
T &=& (t_1 + t_2)/2  \, , \quad t = t_1 - t_2
\label{eq127}
\end{eqnarray}
and write then
\begin{eqnarray}
&& \!\!\!\!\!\!
     \phi_{ij} (\bR, T; \br, t)
\label{eq129}\\
&& \qquad
     = \langle v_i (\bR + \br/2, T + t/2) \, w_i (\bR - \br/2, T - t/2) \rangle \, .
\nonumber
\end{eqnarray}

Clearly we have
\begin{eqnarray}
&& \!\!\!\!\!\!\!\!\!\!\!\!\!
    \phi_{ij} (\bx_1, t_1 ; \bx_2, t_2)
\nonumber\\
&& = \int\!\!\!\int \int\!\!\!\int
    \langle {\hat{v}}_i (\bk_1, \omega_1) {\hat{w}}_j (\bk_1, \omega_1) \rangle
\label{eq131}\\
&& \qquad \quad
\exp( \iu(\bk_1 \cdot \bx_1 + \bk_2 \cdot \bx_2 - \omega_1 t_1  - \omega_2 t_2 )) \,
\nonumber\\
&& \qquad  \qquad \qquad \qquad \qquad
\dd^3 k_1 \, \dd \omega_1 \, \dd^3 k_2 \, \dd \omega_2 \, .
\nonumber
\end{eqnarray}
In addition to (\ref{eq127}) we introduce
\begin{eqnarray}
\bK &=& \bk_1 + \bk_2 \, , \quad \bk = (\bk_1 - \bk_2) / 2
\nonumber\\
\Omega &=& \omega_1 + \omega_2 \, , \quad \omega = (\omega_1 - \omega_2) / 2
\label{eq133}
\end{eqnarray}
and arrive so at
\begin{eqnarray}
&& \!\!\!\!\!\!\!\!\!\!\!\!\!
   \phi_{ij} (\bR, T; \br, t)
\label{eq135}\\
&& = \int\!\!\!\int {\tilde{\phi}}_{ij} (\bR, T; \bk, \omega)
   \exp{\iu (\bk \cdot \br - \omega t)}
   \dd^3 k \, \dd \omega
\nonumber
\end{eqnarray}
with
\begin{eqnarray}
&& \!\!\!\!\!\!\!\!\!\!\!\!\!
    \tilde{\phi}_{ij} (\bR, T; \bk, \omega)
\nonumber\\
&=& \int\!\!\!\int \langle {\hat{v}}_i (\bk + \bK/2, \omega + \Omega/2)
\label{eq137}\\
&& \qquad \qquad
    {\hat{w}}_j (- \bk + \bK/2, - \omega + \Omega/2) \rangle
\nonumber\\
&& \quad \quad \quad \quad \quad \quad \quad \quad \quad
    \exp(\iu (\bK \cdot \bR - \Omega T)) \,
    \dd^3 K \, \dd \Omega \, .
\nonumber
\end{eqnarray}

In the sense of (\ref{eq129}) we introduce in view of the following calculations
\begin{eqnarray}
&& \chi_{ij} (\bR, T; \br, t)
\nonumber\\
&& \qquad
   =  \langle u_i (\bR + \br/2, T + t/2) \, b_j (\bR - \br/2, T - t/2) \rangle
\nonumber\\
&& v_{ij} (\bR, T; \br, t)
\label{eq139}\\
&& \qquad
   =  \langle u_i (\bR + \br/2, T + t/2) \, u_j (\bR - \br/2, T - t/2) \rangle \, ,
\nonumber
\end{eqnarray}
and denote the quantities that correspond to ${\tilde{\phi}}_{ij}$
{by ${\tilde{\chi}}_{ij}$ and $\tilde{v}_{ij}$, respectively. We
extend these definitions to cases where $u_i$ is replaced by
$u_i^{(\alpha)}$, and $b_j$ or $u_j$ by $b_j^{(\beta)}$ or
$u_j^{(\beta)}$, and use then the notations $\chi_{ij}^{(\alpha
\beta)}$, $v_{ij}^{(\alpha \beta)}$, ${\tilde{\chi}}_{ij}^{(\alpha
\beta)}$ and $\tilde{v}_{ij}^{(\alpha \beta)}$}. For the correlation
tensors $v_{ij}^{(00)}$ and $\tilde{v}_{ij}^{(00)}$ of the
background turbulence we write simply $v_{ij}^{(0)}$ and
$\tilde{v}_{ij}^{(0)}$. Since $\bnab \cdot \bu^{(0)} = 0$ we have
\begin{equation}
k_j {\tilde{v}}_{ji}^{(0)} = \frac{\iu}{2} \nabla_j {\tilde{v}}_{ji}^{(0)} \, , \quad
    k_j {\tilde{v}}_{ij}^{(0)} = - \frac{\iu}{2} \nabla_j {\tilde{v}}_{ij}^{(0)} \, .
\label{eq140}
\end{equation}
{If, as here, both $\bR$ and $\br$ occur in arguments, $\nabla_i$
has to be understood as $\p / \p R_i$.}

\subsection{Preparations for the calculation of $\bscE$}
\label{sec43}

Returning now to the electromotive force $\bscE$ we note first that
\begin{eqnarray}
{\cal{E}}_i (\bR, T) &=& \epsilon_{ilm} \chi_{lm} (\bR, T; 0, 0)
\nonumber\\
&=& \epsilon_{ilm} \int\!\!\!\int {\tilde{\chi}}_{lm} (\bR, T; \bk, \omega) \,
    \dd^3 k \, \dd \omega \, .
\label{eq141}
\end{eqnarray}

Our next goal is to express $\bscE$ by the correlation tensor ${\tilde{v}}_{ij}^{(0)}$.
For this purpose we subject the differential equations for $u_i^{(1)}$, $b_i^{(0)}$
and $b_i^{(1)}$ given by (\ref{eq111}) to a Fourier transformation,
which results in algebraic equations for $\hu_i^{(1)}$, $\hb_i^{(0)}$
and $\hb_i^{(1)}$.
In addition we apply the projection operator
$P_{ij}(\bk) = \delta_{ij} - k_i k_j / k^2$
on that for $\hu_i^{(1)}$.
In this way we obtain
\begin{eqnarray}
&& \hu_i^{(1)} = N (\bk, \omega)
    \big(- U_{ij} \hu_j^{(0)}
\nonumber\\
&& \qquad
    + U_{jk} \big( k_j \frac{\partial \hu_i^{(0)}}{\partial k_k}
    + 2 \frac{k_i k_j}{k^2} \hu_k^{(0)} \big)
    + \Omega_{ij} \hu_j^{(0)} \big)
\nonumber\\
&& \hb_i^{(0)} = E (\bk, \omega)
    \big( \iu (\bk \cdot \bB) \hu_i^{(0)}
    - B_{ij} \hu_j^{(0)}
    - B_{jk} k_j \frac{\partial \hu_i^{(0)}}{\partial k_k} \big)
\nonumber\\
&& \hb_i^{(1)} = E (\bk, \omega)
    \big( \iu (\bk \cdot \bB) \hu_i^{(1)}
\label{eq143}\\
&& \qquad
    - B_{ij} \hu_j^{(1)}
    - B_{jk} k_j \frac{\partial \hu_i^{(1)}}{\partial k_k}
    + U_{ij} \hb_j^{(0)}
    + U_{jk} k_j \frac{\partial \hb_i^{(0)}}{\partial k_k} \big)
\nonumber\\
&& \hat{u}_i^{(0)} k_i \, = \, \hat{u}_i^{(1)} k_i \,
    = \, \hat{b}_i^{(0)} k_i \, = \, \hat{b}_i^{(1)} k_i \, = \, 0
\nonumber
\end{eqnarray}
with the abbreviations $N$, $E$ and $\Omega_{ij}$ defined by
\begin{eqnarray}
N (\bk, \omega) &=& \frac{1}{\nu k^2 - \iu \omega} \, , \quad
    E (\bk, \omega) = \frac{1}{\eta k^2 - \iu \omega}
\nonumber\\
\Omega_{ij} (\bk) &=& 2 \epsilon_{ijk} \frac{(\bk \cdot \bOmega)}{k^2} k_k \, .
\label{eq144}
\end{eqnarray}

\subsection{Calculation of $\bscE^{(0)}$}

We consider now $\bscE$ and the corresponding quantities like $a_{ij}$
and $b_{ijk}$  at $\bR = {\bf 0}$ and $T = 0$.
If we drop the arguments $\bR$ and $T$ we always refer to $\bR = {\bf 0}$ and $T = 0$.
As already mentioned we restrict ourselves on an approximation in which $\bscE$
consists only of the terms $\bscE^{(0)}$ and $\bscE^{(1)}$
in the expansion (\ref{eq109}).

Let us start with $\bscE^{(0)}$.
Clearly $\bscE^{(0)}$ and the corresponding contributions $a_{ij}^{(0)}$
and $b_{ijk}^{(0)}$ to $a_{ij}$ and $b_{ijk}$ are independent
of $\bOmega$, $\bW$ and $\bD$.
In view of $\bscE^{(0)}$ we consider first the contribution
$\chi_{jk}^{(0)}$ to $\chi_{jk}$.
By reasons which will become clear soon we consider for a moment $\chi_{jk}^{(0)} (\bR, T)$
with arbitrary $\bR$ and $T$ and will go only later to the limit $\bR \to {\bf 0}$
and put $T = 0$.
We introduce the notation
\begin{eqnarray}
[f(\bk, \omega)]_+ &=& f(\bk + \bK/2, \omega + \Omega/2)
\nonumber\\
\, [f(\bk, \omega)]_- &=& f(- \bk + \bK/2, - \omega + \Omega/2) \, ,
\label{eq161}
\end{eqnarray}
where $f$ means an arbitrary function.
Then we have
\begin{eqnarray}
&& \!\!\!\!\!\!\!\!\!\!\!\!
\chi_{lm}^{(0)}(\bR, T) = \int\!\!\!\int \int\!\!\!\int \big\langle
     [\hu_l^{(0)}]_+
\nonumber\\
&& \!\!\!\!\!\!\!\!
\quad [\iu B_j E k_j \hu_m^{(0)}
     - B_{jk} \big(E \delta_{jm} \hu_k^{(0)}
     - E k_j \frac{\partial \hu_m^{(0)}}{\partial k_k} \big) ]_- \big\rangle
\label{eq163}\\
&& \qquad \qquad \qquad
     \exp(\iu ((\bK \cdot \bR) - \Omega T))
     \, \dd^3K \, \dd \Omega \, \dd^3k \, \dd \omega \, .
\nonumber
\end{eqnarray}
For the sake of simplicity we have dropped the arguments $\bk$ and $\omega$
of $\hu_i^{(0)}$ and $E$.

For the evaluation of this and similar integrals two relations
are of particular interest.
To explain them we note first that
\begin{eqnarray}
[\frac{\partial f(\bk, \omega)}{\partial k_i}]_+
   &=& \big( \frac{1}{2} \frac{\partial}{\partial k_i}
   + \frac{\partial}{\partial K_i} \big) [f(\bk, \omega)]_+
\nonumber\\
\, [\frac{\partial f(\bk, \omega)}{\partial k_i}]_-
   &=& - \big(\frac{1}{2} \frac{\partial}{\partial k_i}
   - \frac{\partial}{\partial K_i} \big) [f(\bk, \omega{)}]_-
\label{eq167}
\end{eqnarray}
and
\begin{eqnarray}
\big(\frac{1}{2} \frac{\partial}{\partial k_i}
   - \frac{\partial}{\partial K_i} \big) [f(\bk, \omega{)}]_+ &=& 0
\nonumber\\
\big( \frac{1}{2} \frac{\partial}{\partial k_i}
   + \frac{\partial}{\partial K_i} \big) [f(\bk, \omega)]_- &=& 0  \, .
\label{eq169}
\end{eqnarray}
On this basis we find with the help of integrations by parts
\begin{eqnarray}
&& \!\!\!\!\!\!\!
\int\!\!\!\int \int\!\!\!\int [F(\bk, \omega)]_+ \, [G(\bk, \omega)]_- \,
   [\frac{\partial H(\bk, \omega)}{\partial k_i}]_-
\nonumber\\
&& \quad
\exp(\iu ((\bK \cdot \bR) - \Omega T ))
   \, \dd^3K \, \dd \Omega \, \dd^3k \, \dd \omega
\nonumber\\
&& \!\!\!\!\
   = - \int\!\!\!\int \int\!\!\!\int [F(\bk, \omega)]_+
   \, [\frac{\partial G(\bk, \omega)}{\partial k_i}]_- \,[H(\bk, \omega)]_-
\label{eq171}\\
&& \qquad \qquad
   \exp(\iu ((\bK \cdot \bR) - \Omega T{)})
   \, \dd^3K \, \dd \Omega \, \dd^3k \, \dd \omega + O(\bR)
\nonumber
\end{eqnarray}
and an analogous relation with $[ \cdots ]_+$ exchanged by $[ \cdots
]_-$ and vice versa.

Starting from (\ref{eq163}) and using (\ref{eq171}) we find
\begin{eqnarray}
\chi_{lm}^{(0)}(\bR, T) &=& \int\!\!\!\int \int\!\!\!\int \big\{
     \iu B_j [E k_j]_- \, \langle [\hu_l^{(0)}]_+ \, [\hu_m^{(0)}]_- \rangle
\nonumber\\
&& \qquad
     - B_{jk} \big([E]_- \,\delta_{jm} \langle [\hu_l^{(0)}]_+ \, [\hu_k^{(0)}]_- \rangle
\nonumber\\
&& \qquad
     - [\frac{\partial}{\partial k_k} (E k_j)]_-
     \, \langle [\hu_l^{(0)}]_+ \, [\hu_m^{(0)}]_- \rangle \big) \big\}
\label{eq173}\\
&& \!\!\!\!\!\!\!\!\!\!
     \exp(\iu ((\bK \cdot \bR) - \Omega T))
     \, \dd^3K \, \dd \Omega \, \dd^3k \, \dd \omega \, + O(\bR) \, .
\nonumber
\end{eqnarray}
We conclude then
\begin{eqnarray}
&& \!\!\!\!\!\!\!\!\!\!\!\!\!\!
    a_{ij}^{(0)}(\bR, T)
\nonumber\\
&& \!\!\!\!\!\!
    = \iu \epsilon_{ilm} \int\!\!\!\int \int\!\!\!\int
    [E k_j]_- \, \langle [\hu_l^{(0)}]_+ \, [\hu_m^{(0)}]_- \rangle
\label{eq175}\\
&& \exp(\iu ((\bK \cdot \bR) - \Omega T))
     \, \dd^3K \, \dd \Omega \, \dd^3k \, \dd \omega \, + O(\bR)
\nonumber
\end{eqnarray}
and
\begin{eqnarray}
&& \!\!\!\!\!\!\!\!\!\!\!\!\!\!
    b_{ijk}^{(0)}(\bR, T)
\nonumber\\
&& \!\!\!\!\!\!
   = - \epsilon_{ilm} \int\!\!\!\int \int\!\!\!\int \big([E]_- \,
   \delta_{jm}\langle [\hu_l^{(0)}]_+ \, [\hu_k^{(0)}]_- \rangle
\nonumber\\
&& \qquad \qquad \qquad
   - [\frac{\partial}{\partial k_k}(E k_j)]_-
     \langle [\hu_l^{(0)}]_+ \, [\hu_m^{(0)}]_- \rangle \big)
\label{eq177}\\
&& \exp(\iu ((\bK \cdot \bR) - \Omega T))
     \, \dd^3K \, \dd \Omega \, \dd^3k \, \dd \omega \, + O(\bR) \, .
\nonumber
\end{eqnarray}

We assume that all mean quantities vary only weakly with $\bR$ and
not with $T$. In that sense we expand $[E k_j]_-$ in (\ref{eq175})
in a series with respect to $\bK$ but neglect all terms of higher
than first order in $\bK$, and put $\Omega = 0$. The first--order
terms have factors $K_i$ under the integrals, and these correspond
to the application of the operator {$- \iu \nabla_i$ to the function
defined by these integrals without $K_i$. Proceeding} now to the
limit $\bR \to {\bf 0}$ and $T = 0$, writing simply $a_{ij}^{(0)}$
instead of $a_{ij}^{(0)} ({\bf 0}, 0)$ and remembering the
definition of ${\tilde{v}}_{ij}^{(0)}(\bR, T, \bk, \omega)$, we find
\begin{eqnarray}
&& \!\!\!\!\!\!\!\!\!\!\!
    a_{ij}^{(0)} = - \epsilon_{ilm} \int\!\!\!\int
    \big(E^* ( \iu k_j - \frac{1}{2} \nabla_j)
\label{eq181}\\
&& \qquad \qquad \qquad \qquad
    - E^{* \prime} \frac{k_j}{2 k} (\bk \cdot \bnab) \big)
    {\tilde{v}}_{lm}^{(0)} \dd^3 k \, \dd \omega \, .
\nonumber
\end{eqnarray}
Here $E^*$ stands for the complex conjugate of $E(\bk, \omega)$,
which is equal to $E(\bk, - \omega)$.
Note that $E^*$ depends only via $k$ on $\bk$.
For this type of functions we use the notation $f^\prime = \partial f / \partial k$.
Furthermore ${\tilde{v}}_{ij}^{(0)}$ and $\nabla_k
{\tilde{v}}_{ij}^{(0)}$ stands for ${\tilde{v}}_{ij}^{(0)}(\bzo, 0, \bk, \omega)$
and $(\nabla_k {\tilde{v}}_{ij}^{(0)}(\bR, 0, \bk, \omega))_{\bR = \bzo}$, respectively.

Starting from (\ref{eq177}) for $b_{ijk}^{(0)}(\bR, T)$ we proceed analogously.
Since, however, $b_{ijk}$ is connected with the
derivatives of $\bmB$ we replace $[E]_-$ and $[\partial(E k_j) / \partial k_k]_-$
simply by their values at $\bK = {\bf 0}$ and
$\Omega = 0$, that is, ignore any derivatives of
${\tilde{v}}_{ij}^{(0)}$. So we arrive at
\begin{equation}
b_{ijk}^{(0)} = \int\!\!\!\int \big( \epsilon_{ijl} E^*
    {\tilde{v}}_{lk}^{(0)}
    + \epsilon_{ilm} E^{* \prime} \frac{k_j k_k}{k} {\tilde{v}}_{lm}^{(0)} \big)
    \, \dd^3k \, \dd \omega \, .
\label{eq185}
\end{equation}
We have dropped contributions to $b_{ijk}^{(0)}$ proportional to $\delta_{jk}$,
which because of $\bnab \cdot \bmB = 0$ do not contribute to $\bscE$.

The results (\ref{eq181}) and (\ref{eq185}) agree with earlier ones,
e.g., those given in \cite{krauseetal80}.

\subsection{Calculation of $\bscE^{(1)}$}

Let us now consider $\bscE^{(1)}$ and the corresponding
contributions $a_{ij}^{(1)}$ and $b_{ijk}^{(1)}$ to $a_{ij}$ and
$b_{ijk}$. $\bscE^{(1)}$ is a sum of three terms, the first one
linear and homogeneous in $\bOmega$ and the second and third ones
linear and homogeneous in $\bW$ or $\bD$, respectively. {Likewise}
$a_{ij}^{(1)}$ and $b_{ijk}^{(1)}$ are sums of three terms, which
are again linear and homogeneous in $\bOmega$, $\bW$ and $\bD$. We
denote the corresponding contributions to $a_{ij}^{(1)}$ and
$b_{ijk}^{(1)}$ by $a_{ij}^{(\Omega)}$, $a_{ij}^{(W)}$,
$a_{ij}^{(D)}$, $b_{ijk}^{(\Omega)}$, $b_{ijk}^{(W)}$ and
$b_{ijk}^{(D)}$.

We may calculate the latter quantities in the same way as we did
it with $a_{ij}^{(0)}$ and $b_{ijk}^{(0)}$. Unfortunately the
results are rather bulky. Some simplification is possible if we
split $\tilde{v}_{ij}^{(0)}$ into its symmetric and antisymmetric
part,
\begin{equation}
\tilde{v}_{ij}^{(0)} = \tilde{v}_{ij}^{(s)} + \tilde{v}_{ij}^{(a)} \, , \quad
    \tilde{v}_{ij}^{(s)} = \tilde{v}_{ji}^{(s)} \, , \quad
    \tilde{v}_{ij}^{(a)} = - \tilde{v}_{ji}^{(a)}
\label{eq211}
\end{equation}
and assume that the symmetric part is even and the antisymmetric one is odd in $\bk$,
\begin{eqnarray}
\tilde{v}_{ij}^{(s)} (\bk, \omega) &=& \tilde{v}_{ij}^{(s)} (- \bk, \omega)
\nonumber\\
\tilde{v}_{ij}^{(a)} (\bk, \omega) &=& - \tilde{v}_{ij}^{(a)} (- \bk, \omega) \, .
\label{eq213}
\end{eqnarray}
This assumption is true for any homogeneous turbulence
and also for the form of inhomogeneous turbulence which we will consider later.

The results of the calculations for $a_{ij}^{(\Omega)}$ and $b_{ijk}^{(\Omega)}$ read
\begin{eqnarray}
a_{ij}^{(\Omega)} &=& \int\!\!\!\int
   \bigl\{ E^* (N - N^*) \frac{(\bk \cdot \bOmega)}{k^2} k_i \nabla_j \tilde{v}_{ll}^{(s)}
\nonumber\\
&& \!\!\!\!\!\!\!\!\!\!
    - E^* (N + N^*) \big( \frac{(\bk \cdot \bOmega)}{k^2} (k_j \nabla_i
         - 2 \frac{k_i k_j}{k^2}(\bk \cdot \bnab)) \tilde{v}_{ll}^{(s)}
\nonumber\\
&& + \frac{k_i k_j}{k^2}(\bOmega \cdot \bnab)\tilde{v}_{ll}^{(s)}
    - 2 \frac{k_j(\bk \cdot \bOmega)}{k^2} \nabla_l \tilde{v}_{li}^{(s)} \big)
\label{eq215}\\
&& \!\!\!\!\!\!\!\!\!\!
    + (E^{* \prime} (N - N^*) - E^* (N^\prime + N^{* \prime}))
\nonumber\\
&& \qquad \qquad \qquad
   \frac{k_i k_j(\bk \cdot \bOmega)}{k^3} (\bk \cdot \bnab) \tilde{v}_{ll}^{(s)} \bigl\} \,
   \dd^3k \, \dd\omega
\nonumber
\end{eqnarray}
\begin{eqnarray}
b_{ijk}^{(\Omega)} &=& - 2 \int\!\!\!\int \frac{(\bk \cdot
\bOmega)}{k^2} \bigl\{
    E^* (N + N^*) (k_i \tilde{v}_{jk}^{(s)} - k_j \tilde{v}_{ik}^{(s)})
\nonumber\\
&& \qquad \qquad
    + E^* N^* \delta_{ik} k_j \tilde{v}_{ll}^{(s)}
\label{eq221}\\
&& \qquad \qquad
    - E^{* \prime} (N - N^*)
    \frac{k_i k_j k_k}{k} \tilde{v}_{ll}^{(s)} \bigl\} \, \dd^3k \, \dd\omega \, .
\nonumber
\end{eqnarray}
Again $E^*$ stands for the complex conjugate of $E(\bk, \omega)$,
that is for $E(\bk, - \omega)$. Likewise $N$ means $N(\bk,
\omega)$ and $N^*$ its complex conjugate, that is $N(\bk, - \omega)$.
As before ${\tilde{v}}_{ij}^{(0)}$ and $\nabla_m
{\tilde{v}}_{ij}^{(0)}$ mean ${\tilde{v}}_{ij}^{(0)}({\bf 0}, 0,
\bk, \omega)$ and $(\nabla_m {\tilde{v}}_{ij}^{(0)}(\bR, 0, \bk,
\omega))_{\bR = {\bf 0}}$, respectively.
As in the case of $b_{ijk}^{(0)}$ contributions to $b_{ijk}^{(\Omega)}$
with $\delta_{jk}$ have been dropped.

The corresponding results for $a_{ij}^{(W)}$, $a_{ij}^{(D)}$, $b_{ijk}^{(W)}$ and $b_{ijk}^{(D)}$
are given in Appendix~\ref{aijetc}.

\subsection{Results for $\bscE$ with a specific velocity correlation tensor}

We now specify the correlation tensor $\tilde{v}_{ij}^{(0)} (\bR, T,
\bk, \omega)$ so that it corresponds to an inhomogeneous turbulence
deviating from a homogeneous isotropic mirror--symmetric and
statistically steady one only by { a gradient}  of the turbulence
intensity. In that sense we put
\begin{eqnarray}
&& \!\!\!\!
    \tilde{v}_{ij}^{(0)} (\bR, T, \bk, \omega)
\label{eq231}\\
&& \qquad
    = \frac{1}{2} \big(P_{ij}(\bk)
    + \frac{\iu}{2 k^2}(k_i \nabla_j - k_j \nabla_i) \big) W (\bR, T, k, \omega) \, ,
\nonumber
\end{eqnarray}
where again $P_{ij}(\bk) = (\delta_{ij} - k_i k_j / k^2)$.
Here $W (\bR, T, k, \omega)$ is the Fourier transform
of $\langle \bu (\bR + \br/2, T + t/2) \cdot \bu (\bR - \br/2, T - t/2) \rangle$
with respect to $\br$ and $t$,
\begin{eqnarray}
&& \!\!\!\!\!\!
    \int\!\!\!\int W (\bR, T, k, \omega) \, \exp(\iu (\bk \cdot \br - \omega t)) \, \dd^3k \, \dd\omega
\label{eq233}\\
&& \qquad
    = \langle \bu (\bR + \br/2, T + t/2) \cdot \bu (\bR - \br/2, T - t/2) \rangle \, ;
\nonumber
\end{eqnarray}
see also \cite{raedleretal03}.
Note that (\ref{eq231}) satisfies both (\ref{eq140}) and (\ref{eq213}).
Anticipating that we will later specify $W(\bR, T, k, \omega)$ as a product
of a factor $\langle {u^{(0)}}^2 \rangle$ depending on $\bR$ and $T$
and a factor depending on $k$ and $\omega$ only we put
\begin{equation}
\bnab W (\bR, T, k, \omega) = \bg \, W (\bR, T, k, \omega)
\label{eq235}
\end{equation}
and interpret $\bg$
as $\bnab \langle {\bu^{(0)}}^2 \rangle / \langle {\bu^{(0)}}^2 \rangle$.

We now specify the results
for $a_{ij}^{(0)}$, $a_{ij}^{(\Omega)}, \cdots \, b_{ijk}^{(D)}$
given by (\ref{eq181}), (\ref{eq185}), (\ref{eq215}), (\ref{eq221}) and (\ref{eq217}) -- (\ref{eq225})
with the ansatz (\ref{eq231}) for $\tilde{v}_{ij}^{(0)}$.
We further use the relations
\begin{eqnarray}
&& \int k_i k_j f(k) \dd^3k = \frac{1}{3} \, \delta_{ij} \int k^2 f(k) \dd^3k
\nonumber\\
&& \int k_i k_j k_k k_l f(k) \dd^3k
\label{eq237}\\
&& \qquad \qquad
   = \frac{1}{15} \, (\delta_{ij} \delta_{kl} + \delta_{ik} \delta_{jl} + \delta_{il} \delta_{jk})
   \int k^4 f(k) \dd^3k \, ,
\nonumber
\end{eqnarray}
which apply for all functions $f$ depending on $\bk$ only via $k$.
The integrals are over all $\bk$.

In this way we find results for the coefficients $\gamma^{(0)}$, $\beta^{(0)}$, $\alpha_1^{(\Omega)}$,
$\cdots$ $\kappa^{(D)}$, say generally $f$, in the form
\begin{equation}
f = 4 \pi \, \int_{k=0}^\infty \int_{\omega = - \infty}^\infty  \tilde{f} (k, \omega) \,
   W (k, \omega) \, k^2 \, \dd k \, \dd \omega \, .
\label{eq241}
\end{equation}
As for $\gamma^{(0)}$, $\beta^{(0)}$, $\alpha_1^{(\Omega)}$,
$\cdots$ $\kappa^{(\Omega)}$ the $\tilde{f}$ are given by
\begin{eqnarray}
&& 2 {\tilde{\gamma}}^{(0)} = {\tilde{\beta}}^{(0)}
   = \frac{1}{3} \, \frac{\eta k^2}{(\eta k^2)^2 + \o^2}
\nonumber\\
&& {\tilde{\alpha}}_1^{(\Omega)} = \frac{4}{15} \Big(
\frac{\eta \nu k^4 ((\nu k^2)^2 + 3 \o^2)}{((\eta k^2)^2 + \o^2) \,((\nu k^2)^2 + \o^2)^2}
\nonumber\\
&& \qquad \qquad \qquad
   + \frac{2 (\eta k^2)^2 \o^2}{((\eta k^2)^2 + \o^2)^2 \, ((\nu k^2)^2 + \o^2)} \Big)
\nonumber\\
&& {\tilde{\alpha}}_2^{(\Omega)} = - \frac{1}{15} \Big(
   \frac{2 \eta \nu k^4 (3 (\nu k^2)^2 - \o^2)}{((\eta k^2)^2 + \o^2) \, ((\nu k^2)^2 + \o^2)^2}
\label{eq243}\\
&& \qquad \qquad \qquad
    - \frac{(3 (\eta k^2)^2 - 5 \o^2) \o^2}{((\eta k^2)^2 + \o^2)^2 \, ((\nu k^2)^2 + \o^2)} \Big)
\nonumber\\
&& {\tilde{\gamma}}^{(\Omega)} = {\tilde{\delta}}^{(\Omega)}
   = - \frac{1}{3} \, \frac{\omega^2}{((\eta k^2)^2 + \o^2) \, ((\nu k^2)^2 + \o^2)}
\nonumber\\
&& {\tilde{\kappa}}^{(\Omega)} = {\frac{2}{15}} \,
   \frac{(11 (\eta k^2)^2 - 5 \o^2) \o^2}{((\eta k^2)^2 + \o^2)^2 ((\nu k^2)^2 + \o^2)} \, .
\nonumber
\end{eqnarray}
The corresponding results for $\alpha_1^{(W)}$, $\alpha_1^{(W)}$
$\cdots$ $\kappa^{(D)}$ are given in Appendix~\ref{ftilde}. Note
that not only $\gamma^{(0)}$ and $\beta^{(0)}$ are independent of
$\nu$ but also $\delta^{(W)}$. {Whereas this independence is quite
natural for $\gamma^{(0)}$ and $\beta^{(0)}$, it results from an
accidental compensation of contributions in the case of
$\delta^{(W)}$.}

\subsection{Specific results}

Let us now calculate the the coefficients $\gamma^{(0)}$, $\beta^{(0)}$, $\alpha_1^{(\Omega)}$,
$\alpha_2^{(\Omega)}$, $\cdots$ $\kappa^{(D)}$
according to (\ref{eq241}), (\ref{eq243}) and (\ref{eq245}) with a specific ansatz
for $W (\bR, T; \bk, \omega)$, that is for
$\langle \bu (\bR + \br/2, T + t/2) \cdot \bu (\bR - \br/2, T - t/2) \rangle$.
We put
\begin{eqnarray}
&& \!\!\!\!\!\!\!\!\!\!\!
   \langle \bu (\bR + \br/2, T + t/2) \cdot \bu (\bR - \br/2, T - t/2) \rangle
\nonumber\\
&& \qquad \qquad
   = \ol{u^2} (\bR, T) \exp( - r^2 / 2 \lambda^2_c - t / |\tau_c|) \, .
\label{eq251}
\end{eqnarray}
Simplifying the notation we have written $\ol{u^2}$
instead of $\langle {u^{(0)}}^2 \rangle$, that is,
$\ol{u^2}$ describes the turbulence intensity in the limit
of vanishing Coriolis force and mean velocity gradient.
Further $\lambda_c$ and $\tau_c$ are correlation length and time in this limit.
We refrain here from considering $\lambda_c$ and $\tau_c$ as functions of $k$ and $\omega$.
Because of (\ref{eq233}) relation (\ref{eq251}) is equivalent to
\begin{equation}
W = \ol{u^2} (\bR, T) \,  \frac{2 \, \lambda_c^3 \tau_c}{3 \, (2 \pi)^{5/2}} \,
    \frac{(k \lambda_c)^2 \, \exp(- (k \lambda_c)^2/2)}{1 + (\omega \tau_c)^2} \, .
\label{eq253}
\end{equation}

In what follows we use the dimensionless parameters
\begin{equation}
q = \lambda^2_c / \eta \tau_c  \, , \quad
p = \lambda^2_c / \nu \tau_c  \, \quad
P_m = \nu / \eta \, .
\label{eq261}
\end{equation}
The quantity $q$ is the ratio of the magnetic diffusion time $\lambda^2_c / \eta$
to the correlation time $\tau_c$.
We speak simply of low--conductivity limit if $q \to 0$,
and of high--conductivity limit if $q \to \infty$,
knowing that these limits can also be reached with any finite $\eta$
but $\tau_c \to \infty$ or $\tau_c \to 0$, respectively.
Likewise $p$ is the ratio of the hydrodynamic decay time $\lambda^2_c / \nu$
to the correlation time $\tau_c$, and $p \to 0$ and $p \to \infty$
are denoted as the high and low viscosity limits, respectively.
$Pm$ is the magnetic Prandtl number of the fluid,
and it holds $P_m = q / p$.
Furthermore we introduce the magnetic Reynolds number $Rm$, the hydrodynamic Reynolds number $Re$
and the Strouhal number $St$ by
\begin{equation}
Rm = \frac{u_c \lambda_c}{\eta} \, , \quad
Re = \frac{u_c \lambda_c}{\nu} \, , \quad
St = \frac{u_c \tau_c}{\lambda_c} \, ,
\label{eq263}
\end{equation}
where $u_c = \sqrt{\ol{u^2}}$. {For a realistic turbulence $St$ is
close to unity. Then $q$ and $p$ are close to $Rm$ and $Re$,
respectively.}

\begin{figure*}
 \includegraphics[width=\textwidth]{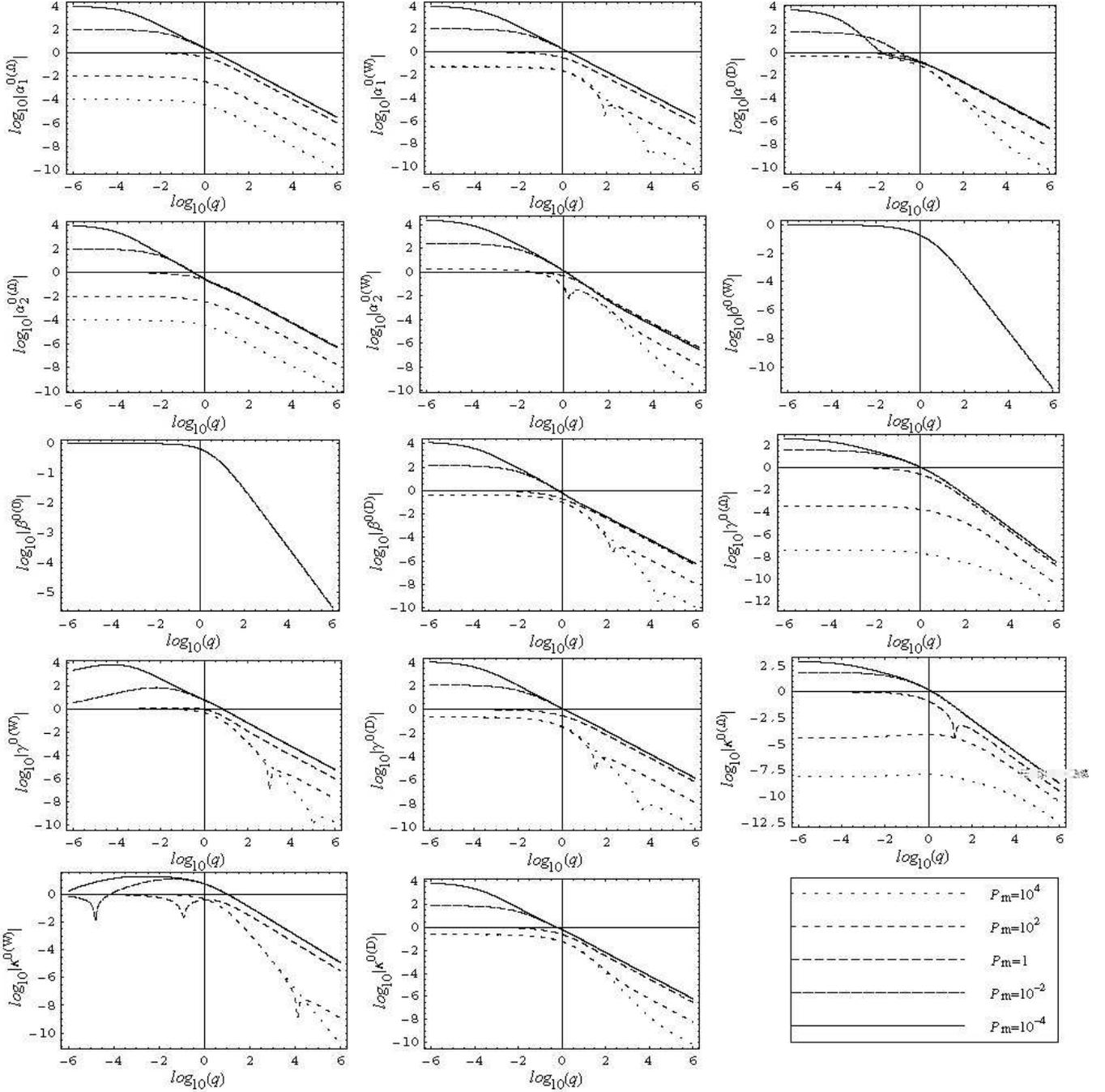}
 \caption{The dependence of  the coefficients $\alpha^{o(\Omega)}_1$,
$\alpha^{o(\Omega)}_1$, $\cdots$ $\kappa^{o(\Omega)}$ on $P_m$ and $q$.
Note that $\gamma^{o(0)}$ coincides with $\beta^{o(0)}$,
and $\delta^{o(\Omega)}$ with $\gamma^{o(\Omega)}$.
The different line styles correspond to different values of
$P_m$, see the last frame.
For all $P_m$ these coefficients are positive as long as $q$ is small.
In some cases the signs change as $q$ grows.
This is indicated by tips of the curves.}
\label{fig1}
\end{figure*}

We return now to the representation (\ref{eq27}) for $\bscE$, again with
$\bg = \bnab \ol{u^2} / \ol{u^2}$.
We give our results for the coefficients in this representation
first in a form suitable for application to the dynamo experiment mentioned above,
where $q$ is at least not large compared to unity.
This form reads
\begin{eqnarray}
\alpha^{(\Omega)}_1 &=& (4 /45) \, Rm^2 \, \lambda_c^2 \, \alpha^{o(\Omega)}_1 (P_m, q)
\nonumber\\
\alpha^{(\Omega)}_2&=& - (2 / 15) \, Rm^2 \, \lambda_c^2 \, \alpha^{o(\Omega)}_2 (P_m, q)
\nonumber\\
\alpha^{(W)}_1 &=& (19 / 360) \, Rm^2 \, \lambda_c^2  \, \alpha^{o(W)}_1 (P_m, q)
\nonumber\\
\alpha^{(W)}_2&=& - (7 / 720{)} \, Rm^2 \, \lambda_c^2  \,
\alpha^{o(W)}_2 (P_m, q)
\nonumber\\
\alpha^{(D)} &=& - (7 / 120) \, Rm^2 \, \lambda_c^2 \, \alpha^{o(D)} (P_m, q)
\label{eq267a}\\\gamma^{(0)} &=& \frac{1}{18} \, Rm^2 \, \eta \, \gamma^{o(0)}(q)
\nonumber\\
\gamma^{(\Omega)}&=& - (\sqrt{\pi} / 36 \sqrt{2}) \, Rm^2 \, \lambda_c^2 \,
\sqrt{q} \, \gamma^{o(\Omega)} (P_m, q)
\nonumber\\
\gamma^{(W)} &=& - (1 / 144) \, Rm^2 \, \lambda_c^2 \,\gamma^{o(W)} (P_m, q)
\nonumber\\
\gamma^{(D)}&=& - (13 / 120) \, Rm^2 \, \lambda_c^2  \, \gamma^{o(D)} (P_m, q)
\nonumber
\end{eqnarray}
\begin{eqnarray}
\beta^{(0)} &=& (1 / 9) Rm^2 \eta \beta^{o(0)} (q)
\nonumber\\
\beta^{(D)}&=& (7 / 90) \, Rm^2 \, \lambda_c^2 \, \beta^{o(D)} (P_m, q)
\nonumber\\
\delta^{(\Omega)} &=& - (\sqrt{\pi} / 36 \sqrt{2}) \, Rm^2 \, \lambda_c^2 \,
\sqrt{q} \, \delta^{o(\Omega)} (P_m, q)
\nonumber\\
\delta^{(W)}&=& (1 / 36) \, Rm^2 \, \lambda_c^2 \, \delta^{o(W)} (q)
\label{eq267b}\\
\kappa^{(\Omega)} &=& (\sqrt{\pi} / {18} \sqrt{2}) \, Rm^2 \,
\lambda_c^2 \, \sqrt{q} \kappa^{o(\Omega)} (P_m, q)
\nonumber\\
\kappa^{(W)}&=& - (1 / {90}) \, Rm^2 \, \lambda_c^2  \,
\kappa^{o(W)} (P_m, q)
\nonumber\\
\kappa^{(D)} &=& (13 / 90) \, Rm^2 \, \lambda_c^2 \, \kappa^{o(D)} (P_m, q) \, .
\nonumber
\end{eqnarray}
The numerical factors are chosen such that the functions
$\alpha^{o(\Omega)}_1$, $\alpha^{o(\Omega)}_1$, $\cdots$
$\kappa^{o(\Omega)}$ with $P_m = 1$ approach unity in the
low--conductivity limit $q \to 0$. According to (\ref{eq241}) and
(\ref{eq243}) we have $\gamma^{o(0)} = \beta^{o(0)}$ and
$\gamma^{o(\Omega)} = \delta^{o(\Omega)}$. Figure \ref{fig1} shows
the dependence of the functions $\alpha^{o(\Omega)}_1$,
$\alpha^{o(\Omega)}_1$, $\cdots$ $\kappa^{o(\Omega)}$ on $P_m$ and
$q$.

\begin{figure*}
 \includegraphics[width=\textwidth]{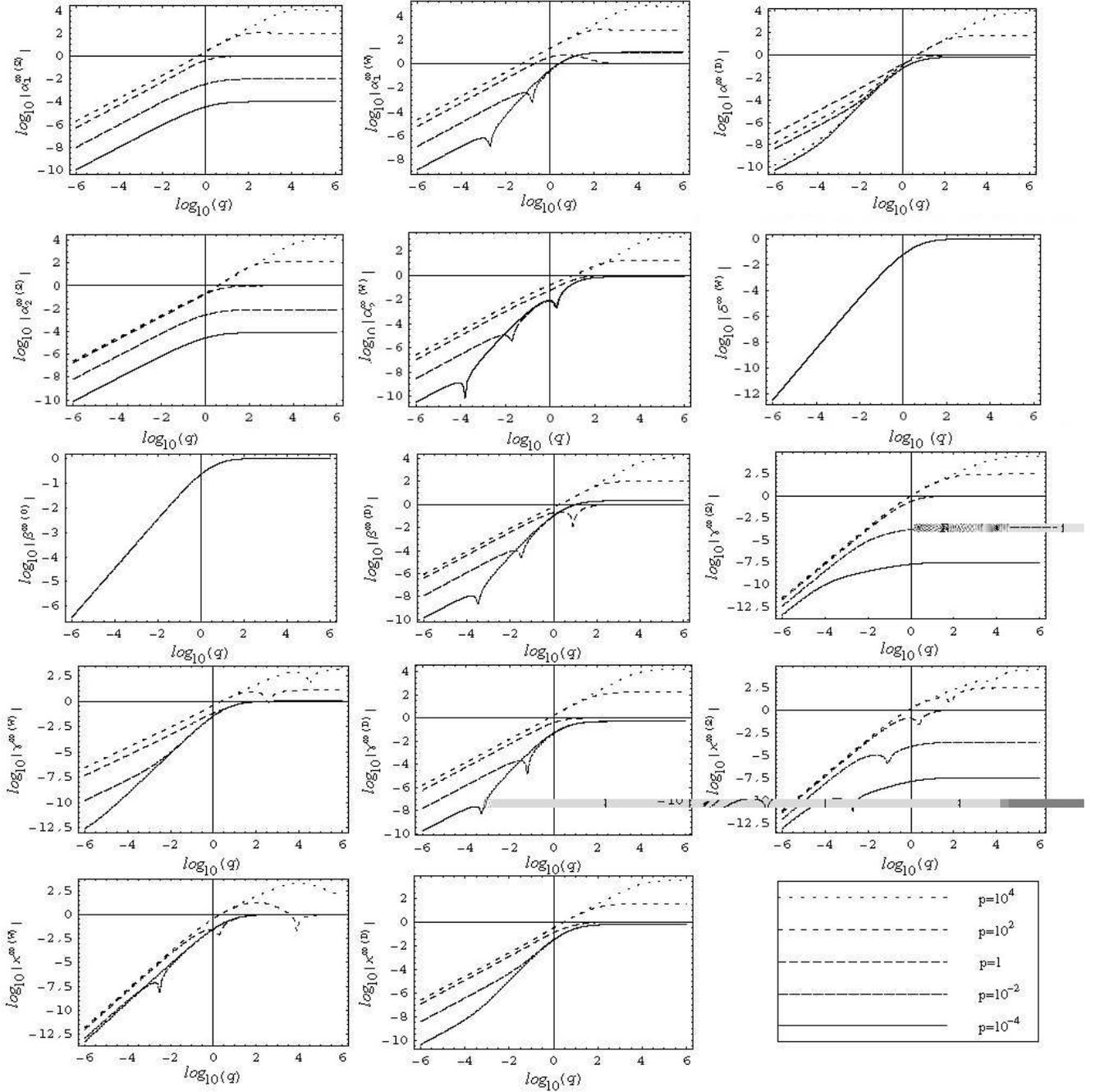}
 \caption{The dependence of  the coefficients $\alpha^{\infty(\Omega)}_1$,
$\alpha^{\infty(\Omega)}_1$, $\cdots$ $\kappa^{\infty(\Omega)}$ on
$p$ and $q$.
Note that $\gamma^{\infty(0)}$ coincides with $\beta^{\infty(0)}$,
and $\delta^{\infty(\Omega)}$ with $\gamma^{\infty(\Omega)}$.
The explanations given with Fig.~\ref{fig1} apply analogously but
for all $p$ these coefficients are positive as long as $q$ is large.}
\label{fig2}
\end{figure*}

In astrophysical applications the high--conductivity limit $q \to \infty$
is of particular interest.
Then a modified representation of these results seems appropriate,
\begin{eqnarray}
\alpha^{(\Omega)}_1 &=& (4 / 45) \, \ol{u^2} \tau_c^2 \, \alpha^{\infty(\Omega)}_1 (p, q)
\nonumber\\
\alpha^{(\Omega)}_2&=& - (1 / 90) (22 - 5 \xi) \, \ol{u^2} \tau_c^2 \, \alpha^{\infty(\Omega)}_2 (p, q)
\nonumber\\
\alpha^{(W)}_1 &=& (1 / 72) (\xi - 1) \, \ol{u^2} \tau_c^2 \, \alpha^{\infty(W)}_1 (p, q)
\nonumber\\
\alpha^{(W)}_2&=& - (1 / 144) (11 + \xi) \, \ol{u^2} \tau_c^2 \, \alpha^{\infty(W)}_2 (p, q)
\nonumber\\
\alpha^{(D)} &=& - (1 / 360) (29 - 5 \xi) \, \ol{u^2} \tau_c^2 \, \alpha^{\infty(D)} (p, q)
\label{eq271a}\\
\gamma^{(0)} &=& (1 / 6) \, \ol{u^2} \tau_c \, \gamma^{\infty(0)}(q)
\nonumber\\
\gamma^{(\Omega)}&=& - (1 / 18) (2 - \xi) \, \ol{u^2} \tau_c^2 \, \gamma^{\infty(\Omega)} (p, q)
\nonumber\\
\gamma^{(W)} &=& - (1 / 144) (13 + \xi )) \, \ol{u^2} \tau_c^2 \, \gamma^{\infty(W)} (p, q)
\nonumber\\
\gamma^{(D)}&=& - (1 / 72) (7 - \xi) \, \ol{u^2} \tau_c^2 \, \gamma^{\infty(D)} (p, q)
\nonumber
\end{eqnarray}
\begin{eqnarray}
\beta^{(0)} &=& (1 / 3) \, \ol{u^2} \, \tau_c \, \beta^{\infty(0)} (q)
\nonumber\\
\beta^{(D)}&=& - (7 / 90)  \, \ol{u^2} \tau_c^2 \, \beta^{\infty(D)} (p, q)
\nonumber\\
\delta^{(\Omega)} &=& - (1 / 18) (2 - \xi) \, \ol{u^2} \tau_c^2 \, \delta^{\infty(\Omega)} (p, q)
\nonumber\\
\delta^{(W)}&=& (1 / 12) \, \ol{u^2} \tau_c^2 \, \delta^{\infty(W)} (q)
\label{eq271b}\\
\kappa^{(\Omega)} &=& - (1 / {9}) (2 - \xi) \, \ol{u^2} \tau_c^2 \,
\kappa^{\infty(\Omega)} (p, q)
\nonumber\\
\kappa^{(W)}&=& - (1 / {6}) \, \ol{u^2} \tau_c^2 \,
\kappa^{\infty(W)} (p, q)
\nonumber\\
\kappa^{(D)} &=& (23 / 90) \, \ol{u^2} \tau_c^2 \, \kappa^{\infty(D)} (p, q) \, ,
\nonumber
\end{eqnarray}
where $\xi=\sqrt{2 e}(\sqrt{\pi}-2\int_0^{\sqrt{2}}\exp(-t^2)dt)
\approx 1.31$. The functions $\alpha^{\infty(\Omega)}_1$,
$\alpha^{\infty(\Omega)}_1$, $\cdots$ $\kappa^{\infty(\Omega)}$ are
defined such that their values at $p=1$ approach unity as $q \to
\infty$. Note that $\ol{u^2} \tau_c^2 = St^2 \lambda_c^2$. According
to (\ref{eq241}) and (\ref{eq243}) we have now $\gamma^{\infty(0)} =
\beta^{\infty(0)}$ and $\gamma^{\infty(\Omega)} =
\delta^{\infty(\Omega)}$. The functions $\alpha^{\infty(\Omega)}_1$,
$\alpha^{\infty(\Omega)}_1$, $\cdots$ $\kappa^{\infty(\Omega)}$ are
shown in Figure \ref{fig2}.

\section{Discussion}
\label{disc}

\subsection{Assumptions and approximations}
\label{subsecass}

Our results have been gained with some assumptions and
approximations. As usual it has been generally assumed that
electromotive force $\bscE$ depends in the linear and homogeneous
form (\ref{eq07}) on $\bmB$. The only additional assumption
introduced in Section 3, just for the sake of simplicity, is the
linearity of $\bscE$ in the angular velocity $\bOmega$ and the
gradient $\bnab \bmU$ of the mean velocity, that is, some smallness
of the Coriolis force and the shear in the mean motion. In Section
\ref{sec4}, however, some kind of second--order correlation
approximation (SOCA) has been introduced. As long as only results
are considered which are independent of $\bOmega$ and $\bnab \bmU$,
our procedure corresponds to the traditional second--order
correlation approximation; see, e.g., \cite{krauseetal80}. In the
low--conductivity limit, $q \to 0$, a sufficient condition for the
validity of these results is $Rm \ll 1$. In the high-conductivity
limit, $q \to \infty$, the corresponding condition reads $St \ll 1$.
If non--zero $\bOmega$ and $\bnab \bmU$ are taken into account,
additional conditions expressing the smallness of their effects on
the fluid motion have to be satisfied. These conditions are roughly
described {in}  Section \ref{sec41}.

\subsection{Former results}
\label{subsecform}
There is a series of former results for situations covered by our assumptions.
We refer in particular to those in the early works by Steenbeeck et al. \cite{steenbecketal66},
Krause et al. \cite{krauseetal71b}, R\"adler \cite{raedler69},
further to those by Vainshtein et al. \cite{vainshteinetal83}, R\"udiger et al. \cite{ruedigeretal93c}
and Kichatinov et al. \cite{kitchatinovetal94b}.
As far as these results are given in a form that allows a detailed comparison
our results are in satisfying agreement with most of them.
We note that in the calculations by Steenbeck et al. \cite{steenbecketal66},
which revealed the $\alpha$--effect, due to an incorrect assumption
on the velocity correlation tensor, the $\bOmega \x \bJ$--effect does not occur.
The latter was found only later \cite{raedler69}.

We also point out the recent papers by R\"adler, Kleeorin and
Rogachevskii \cite{raedleretal03} (referred to as RKR03 in the
following) and by Rogachevskii and Kleeorin
\cite{rogachevskiietal03} (referred to as RK03). In both of them an
approach is used, which is aimed to go beyond the second--order
correlation approximation by taking into account higher--order
correlations of $\bu$ and $\bb$ at least in some crude way. {It was
suggested by the $\tau$--approximation of turbulence theory and is
therefore called ``$\tau$--approach" in the following.}
Unfortunately, there is no parameter range in which it completely
reproduces the results for the mean--field coefficients obtained
with the second--order correlation approximation; see
\cite{raedleretal05c}. Possibly the assumptions of the
$\tau$--approach, which rely on a developed turbulence with high
hydrodynamic and magnetic Reynolds numbers, $Re$ and $Rm$, exclude
the assumptions used in the second--order correlation approach.
Nevertheless some of the findings of the $\tau$--approach are of
interest for the following.

\subsection{New findings concerning the $\alpha$, $\gamma$ and $\beta$--effects}
\label{subsecalpha}

It is well known that an inhomogeneous turbulence at a rotating body
gives rise to an $\alpha$--effect. In this case the essential
construction elements of the tensor $\balpha$ are the vectors $\bg$
and $\bOmega$ describing the gradient in the turbulence intensity
and the Coriolis force. Our results show in agreement with those by
RK03 that even in the absence of a Coriolis force the combination of
inhomogeneous turbulence, that is non-zero $\bg$, and a gradient of
the mean velocity, $\bnab \bmU$, leads to an $\alpha$--effect. This
is perhaps less surprising if the gradient of the mean velocity
corresponds to a rotation. Then the role of $\bOmega$ in the tensor
$\balpha$ is played by $\bW$. It is however quite remarkable that,
again in combination with inhomogeneous turbulence, also the
symmetric part of the mean velocity gradient, $\bD$, which
corresponds to a deformation, leads to an $\alpha$--effect. This
contribution to $\balpha$ has however some peculiarity, in
particular its trace is equal to {zero; see also
\cite{raedleretal06b}}.

In all models of $\alpha^2$ or $\alpha \omega$ dynamos considered so far
the contributions to the $\alpha$--effect depending on the shear of the mean flow
have been ignored.
It remains to be investigated how they modify the behavior of such dynamos,
in particular that of an $\alpha \omega$ dynamo in the case of very strong differential rotation.

It is also known that the $\gamma$--effect, which describes a transport of mean magnetic flux
and occurs primarily as a consequence of a gradient of the turbulence intensity,
is modified by the Coriolis force, that is, the vector $\bgamma$ contains a part with $\bOmega$.
Our results show in agreement with RK03 that $\bgamma$ possesses also
contributions with both parts of the mean velocity gradient $\bnab \bmU$,
that is, with $\bW$ and $\bD$.

In mean--field electrodynamics instead of the molecular magnetic diffusivity $\eta$
the mean--field diffusivity $\eta + \beta^{(0)}$ occurs.
More generally spoken, the tensor $\bbeta$ has to be added
to the isotropic molecular diffusivity tensor.
It is clear from simple symmetry considerations and can also be seen
in RKR03 and in RK03 that there are no contributions to $\bbeta$
depending on $\bOmega$ or $\bW$
as long as we restrict ourselves to linearity in these quantities.
We have found however, again in agreement with RK03, that there is a contribution
proportional to the symmetric part of the mean velocity gradient $\bnab \bmU$,
that is to $\bD$.
The mean--field diffusivity, and so the mean--field conductivity,
becomes anisotropic as a consequence of the deforming mean motion
described by $\bD$.

Since $\beta^{(0)}$ is always positive it raises the threshold of a dynamo.
Interestingly enough the mean--field diffusivity tensor need not to be positive definite,
and the $\beta$--effect may then well support a dynamo, see \cite{raedleretal06b}.

\subsection{New findings concerning the $\delta$ and $\kappa$--effects}
\label{subsecdelta}

Proceeding to the $\delta$ and $\kappa$--effects we mention first
that already in the case of a homogeneous turbulence at a rotating
body, that is, subject to the Coriolis force, contributions to the
mean electromotive force proportional to $\bOmega \times (\bnab
\times \bmB)$ and to $\bOmega \circ (\bnab \bmB)^{(s)}$ proved to be
possible. {They usually occur simultaneously. As already mentioned
the occurrence of the first one is often referred to as $\bOmega
\times \bJ$--effect. We note that
\begin{eqnarray}
&& \!\!\!\!\!\!\!\!\!\!\!\!\!\!\!
\delta^{(\Omega)} \bOmega \x (\bnab \x \bmB) + \kappa^{(\Omega)} \bOmega \circ (\bnab \bmB)^{(s)}
\nonumber\\
&&  \qquad
   = \zeta_1^{(\Omega)} \, (\bOmega \cdot \bnab) \bmB + \zeta_2^{(\Omega)} \, \bnab (\bOmega \cdot \bmB) \, ,
\label{eq281}
\end{eqnarray}
where
\begin{equation}
\zeta_1^{(\Omega)} = - \delta^{(\Omega)} + \frac{1}{2} \, \kappa^{(\Omega)} \, , \quad
    \zeta_2^{(\Omega)} = \delta^{(\Omega)} + \frac{1}{2} \, \kappa^{(\Omega)} \, .
\label{eq282}
\end{equation}
As long as $\zeta_2^{(\Omega)}$ is independent of position the last
term on the right--hand side is without interest for the induction
equation. Then the $\delta^{(\Omega)}$ and
$\kappa^{(\Omega)}$--effects act, apart from the signs, in the same
way. Interestingly enough, $\zeta_1^{(\Omega)}$ vanishes in both
limits $q \to 0$ and $q \to \infty$. As long as the ansatz
(\ref{eq251}) is adopted and therefore (\ref{eq267b}) and
(\ref{eq271b}) apply, this can easily be seen for $P_m =1$ and $q
\to 0$ from (\ref{eq267b}), and for $p = 1$ and $q \to \infty$ from
(\ref{eq271b}). A more general proof of the above statement on
$\zeta_1^{(\Omega)}$ is given in Appendix \ref{omegaxj}.}

{Let us have a look on the results of the $\tau$--approach for
$\delta^{(\Omega)}$ and $\kappa^{(\Omega)}$ given in RKR03. It seems
plausible to interpret them as results for $q \to \infty$. The
quantity $\zeta_1^{(\Omega)}$ calculated from them is equal to zero
if the correlation time $\tau_{\mathrm{c}}$ is considered as a
constant, but it deviates from zero as soon as its Fourier transform
depends on $k$. This is in conflict with the general result
explained in Appendix \ref{omegaxj}.}

{We recall that the $\delta^{(\Omega)}$--effect, even in the absence
of any $\alpha$--effect, but in combination with differential
rotation, is capable of dynamo action, see
\cite{raedler69b,raedler70,raedler86,roberts72,moffattetal82} and
RKR03. Dynamos of that kind are often labelled as $\bOmega \x
\bJ$--dynamos. Strictly speaking, both the $\delta^{(\Omega)}$ and
the $\kappa^{(\Omega)}$--effects may constitute this dynamo
mechanism if only $\zeta_1^{(\Omega)}$ is non--zero. As a
consequence of the differential rotation, also induction effects
connected with $\bW$ and $\bD$ necessarily play some part in
$\bOmega \x \bJ$--dynamos but have not been considered so far.}

Our above results show that besides the $\bOmega \times \bJ$--effect
also an analogous $\bW \times \bJ$--effect exists, which occurs even
in the absence of the Coriolis force. This effect and the related
ones have already been considered by Urpin \cite{urpin99,urpin99c}
and extensively studied in RK03. However, details of the results by
Urpin seem to be incorrect, and those of RK03 do not agree with
ours, which is a consequence of the fact that the {$\tau$--approach
was used instead of the second--order correlation approximation.
Analogous to (\ref{eq281}) we have
\begin{eqnarray}
&& \!\!\!\!\!\!\!\!\!\!\!\!\!\!\!
    \delta^{(W)} \bW \x (\bnab \x \bmB) + \kappa^{(W)} \bW \circ (\bnab \bmB)^{(s)}
\nonumber\\
&& \qquad
     = \zeta_1^{(W)} \, (\bW \cdot \bnab) \bmB + \zeta_2^{(W)} \, \bW \circ (\bnab \bmB) \, ,
\label{eq283}
\end{eqnarray}
where
\begin{equation}
\zeta_1^{(W)} = - \delta^{(W)} + \frac{1}{2} \, \kappa^{(W)} \, , \quad
     \zeta_2^{(W)} = \delta^{(W)} + \frac{1}{2} \, \kappa^{(W)} \, .
\label{eq284}
\end{equation}
Here $\bW \circ (\bnab \bmB)$ is defined by $(\bW \circ (\bnab
\bmB))_i = W_j \p \mB_j / \p x_i$. For constant $\bW$ it is again a
gradient. If then in addition $\zeta_2^{(W)}$ is independent of
position the $\delta^{(W)}$ and $\kappa^{(W)}$--effects act again in
the same way. In contrast to $\zeta_1^{(\Omega)}$ the coefficient
$\zeta_1^{(W)}$ takes in general non--zero values as $q \to 0$ or $q
\to \infty$.}

Different from the situation with the $\delta^{(\Omega)}$ and $\kappa^{(\Omega)}$--effects,
the $\delta^{(W)}$ and $\kappa^{(W)}$--effects are accompanied
by the $\beta^{(D)}$ and $\kappa^{(D)}$--effects.
Apart from the case in which $\bmU$ corresponds to a rigid--body rotation,
together with $\bW$ also $\bD$ is unequal to zero so that
the $\beta^{(D)}$ and $\kappa^{(D)}$--effects indeed occur.
This makes the comparison between the effects working with $\bOmega$
and those working with $\bW$ more complex.
Note that in contrast to the signs of $\alpha_1^{(\Omega)}$ and $\alpha_1^{(W)}$,
of $\alpha_2^{(\Omega)}$ and $\alpha_2^{(W)}$ and of $\gamma^{(\Omega)}$ and $\gamma^{(W)}$,
those of $\delta^{(\Omega)}$ and $\delta^{(W)}$ differ;
with $\kappa^{(\Omega)}$ and $\kappa^{(W)}$ the situation depends on $q$.

Analogously to the $\bOmega \x \bJ$--dynamo an $\bW \x \bJ$--dynamo
was proposed in RK03, working with the induction effects of
turbulence discussed here, which are due to a mean shear, and the
induction effect due to the shear alone. In a simple model in
Cartesian geometry, using results for $\delta^{(W)}$,
$\kappa^{(W)}$, $\beta^{(D)}$ and $\kappa^{(D)}$ obtained {in the
$\tau$--approach, indeed growing $\bmB$ were found}.  Recently
R\"udiger et al. \cite{ruedigeretal06} pointed out that this model
does not work as a dynamo with $\delta^{(W)}$, $\kappa^{(W)}$,
$\beta^{(D)}$ and $\kappa^{(D)}$ as found in the second--order
correlation approximation. Our consideration in Appendix~\ref{wxj}
confirms this finding. We stress that our negative conclusion apply
only to a simple model of the $\bW \x \bJ$ dynamo and to the range
of validity of the second--order correlation approximation. It
remains to be checked whether this applies to other models, too. In
cylindrical or spherical geometry the $\bOmega \x \bJ$ and $\bW \x
\bJ$ effects occur always simultaneously. The question of a pure
$\bW \x \bJ$ dynamo does not appear.

\appendix

\section{Relations for $a_{ij}^{(W)}$, $a_{ij}^{(D)}$, $b_{ijk}^{(W)}$ and $b_{ijk}^{(D)}$}
\label{aijetc}

Analogous to the results (\ref{eq215}) and (\ref{eq221})
for $a_{ij}^{(\Omega)}$ and $b_{ijk}^{(\Omega)}$ we find
\begin{widetext}
\begin{eqnarray}
a_{ij}^{(W)} &=&  \frac{\iu}{2}\int\!\!\!\int \bigl\{
   E^*(N + N^*) k_j W_l \tilde{v}_{li}^{(a)}
   + 2 E^* N^*  k_i W_l \tilde{v}_{lj}^{(a)}
   + {E^*}^2 k_j W_l \tilde{v}_{li}^{(a)} \big\} \, \dd^3k \, \dd\omega
\nonumber\\
&& + \frac{1}{4} \int\!\!\!\int \bigl\{
    E^* N \big(- W_i \nabla_j \tilde{v}_{ll}^{(s)}
    + W_l \nabla_j \tilde{v}_{li}^{(s)}
    + 4 \frac{k_j (\bW \cdot \bk)}{k^2} \nabla_l \tilde{v}_{li}^{(s)}
    + 2 \frac{k_i (\bW \cdot \bk)}{k^2} \nabla_j \tilde{v}_{ll}^{(s)}
\nonumber\\
&& \qquad \qquad \qquad
    - 2 \frac{k_i k_j}{k^2} (\bW \cdot \bnab) \tilde{v}_{ll}^{(s)}
    + 4 \frac{k_i k_j (\bW \cdot \bk)}{k^4}
    (\bk \cdot \bnab)\ \tilde{v}_{ll}^{(s)}
    - 2 \frac{k_j (\bW \cdot \bk)}{k^2} \nabla_i \tilde{v}_{ll}^{(s)} \big)
\nonumber\\
&& \quad \quad
    + E^* N^* \big(W_i (\nabla_j \tilde{v}_{ll}^{(s)}
    + 2 \nabla_l \tilde{v}_{lj}^{(s)})
    - W_l \nabla_j \tilde{v}_{li}^{(s)}
    - 2 \delta_{ij} W_l \nabla_n \tilde{v}_{ln}^{(s)}
    - 2 \frac{k_i k_j}{k^2} (\bW \cdot \bnab) \tilde{v}_{ll}^{(s)}
\label{eq217}\\
&& \qquad \qquad \qquad
    + 4 \frac{k_i k_j (\bW \cdot \bk)}{k^4}(\bk \cdot \bnab) \tilde{v}_{ll}^{(s)}
    - 2 \frac{k_i (\bW \cdot \bk)}{k^2} \nabla_j \tilde{v}_{ll}^{(s)}
    + 4 \frac{k_j (\bW \cdot \bk)}{k^2} \nabla_l \tilde{v}_{li}^{(s)}
    - 2 \frac{k_j (\bW \cdot \bk)}{k^2} \nabla_i \tilde{v}_{ll}^{(s)} \big)
\nonumber\\
&& \quad \quad
    - (E^{* \prime} N - E^* N^\prime - (E^* N^*)^\prime )
    \frac{k_j}{k} (\bk \cdot \bnab) \big(W_i \tilde{v}_{ll}^{(s)}
    - W_l \tilde{v}_{li}^{(s)}
    - 2 \frac{k_i (\bW \cdot \bk)}{k^2} \tilde{v}_{ll}^{(s)} \big)
\nonumber\\
&& \quad \quad
    + ({E^*}^2 \nabla_j + ({E^*}^2)^\prime \frac{k_j}{k} (\bk \cdot \bnab) )
    (W_i \tilde{v}_{ll}^{(s)} - W_l \tilde{v}_{li}^{(s)})
    \bigl\}  \, \dd^3k \, \dd\omega
\nonumber
\end{eqnarray}
\begin{eqnarray}
a_{ij}^{(D)} &=& \iu \epsilon_{ilm} \int\!\!\!\int \bigl\{
     E^* (N + N^*) k_j (D_{mn} - 2 \frac{k_m k_p}{k^2} D_{pn})
     \tilde{v}_{ln}^{(a)}
     + E^* N^* k_n D_{nj} \tilde{v}_{lm}^{(a)}
\nonumber\\
&& \quad \quad + ( E^* N^\prime + (E^* N^*)^\prime +
\frac{1}{2}({E^*}^2)^\prime )
     \frac{k_j k_p k_n}{k} D_{pn} \tilde{v}_{lm}^{(a)}
     - {E^*}^2 k_j D_{mn} \tilde{v}_{ln}^{(a)}
    \bigl\} \, \dd^3k \, \dd\omega
\nonumber\\
&& - \frac{1}{2} \epsilon_{ilm} \int\!\!\!\int \bigl\{
     E^* (N -N^*) (D_{ln} - 2 \frac{k_l k_p}{k^2} D_{pn})
        \nabla_j \tilde{v}_{mn}^{(s)}
\label{eq219}\\
&& \quad \quad
     + 2 E^*(N + N^*) k_j (\frac{k_p}{k^2} D_{pn} \nabla_l
     + \frac{k_l}{k^2} D_{pn} \nabla_p
     - 2 \frac{k_l k_p}{k^4} D_{pn} (\bk \cdot \bnab)) \tilde{v}_{mn}^{(s)}
\nonumber\\
&& \quad \quad + (E^{* \prime} N - E^* N^\prime - (E^* N^*)^\prime)
     \frac{k_j}{k}(D_{ln} - 2 \frac{k_l k_p}{k^2} D_{pn})
     (\bk \cdot \bnab) \tilde{v}_{mn}^{(s)}
\nonumber\\
&& \quad \quad - {E^*}^2 D_{mn} \nabla_j
     \tilde{v}_{ln}^{(s)}
      - ({E^*}^2)^\prime \frac{k_j}{k} D_{mn} (\bk \cdot \bnab) \tilde{v}_{ln}^{(s)}
     \bigl\} \, \dd^3k \, \dd\omega
\nonumber
\end{eqnarray}
\begin{eqnarray}
b_{ijk}^{(W)} &=& \frac{1}{2} \int\!\!\!\int
    \bigl\{ E^* (N - N^*) (W_i \tilde{v}_{jk}^{(s)} - W_j \tilde{v}_{ik}^{(s)})
    - 2 E^* N \frac{(\bk \cdot \bW)}{k^2}(k_i \tilde{v}_{jk}^{(s)}
    - k_j \tilde{v}_{ik}^{(s)})
\nonumber\\
&& \quad \quad - E^* N^* \big(\delta_{ik}(W_j \tilde{v}_{ll}^{(s)}
    - W_l \tilde{v}_{lj}^{(s)})
    - 2 \big(\frac{k_i k_k}{k^2}(W_j \tilde{v}_{ll}^{(s)} - W_l \tilde{v}_{lj}^{(s)})
    -  \frac{k_j k_k}{k^2}(W_i \tilde{v}_{ll}^{(s)} - W_l \tilde{v}_{li}^{(s)})) \big) \big)
\nonumber\\
&& \quad \quad
    - E^{* \prime} (N - N^* )\frac{k_j k_k}{k}
    \big(W_i \tilde{v}_{ll}^{(s)} - W_l \tilde{v}_{li}^{(s)}
    - 2 \frac{(\bk \cdot \bW)}{k^2} k_i \tilde{v}_{ll}^{(s)} \big)
\label{eq223}\\
&& \quad \quad
    - {E^*}^2 (W_i \tilde{v}_{jk}^{(s)} - \delta_{ij} W_l \tilde{v}_{lk}^{(s)})
    + ({E^*}^2)^\prime \frac{k_j k_k}{k}
    (W_i \tilde{v}_{ll}^{(s)} - W_l \tilde{v}_{li}^{(s)})
    \bigl\} \, \dd^3k \, \dd\omega
\nonumber
\end{eqnarray}
\begin{eqnarray}
b_{ijk}^{(D)} &=& - \int\!\!\!\int \bigl\{
    E^* N \epsilon_{ijl} \big( D_{lm} - 2 \frac{k_l k_n}{k^2} D_{nm} \big) \tilde{v}_{mk}^{(s)}
    + E^* N^* \epsilon_{ijl} \big( D_{km} - 2 \frac{k_k k_n}{k^2} D_{nm} \big) \tilde{v}_{ml}^{(s)}
\nonumber\\
&& \quad \quad
- E^{* \prime} (N - N^*) \epsilon_{ilm}
    \frac{k_j k_k}{k} \big( D_{mn}
    - 2 \frac{k_m k_p}{k^2} D_{pn} \big) \tilde{v}_{nl}^{(s)}
+ (E^* N^\prime + (E^* N^*)^\prime ) \epsilon_{ijl} \frac{k_m k_n}{k} D_{mn}
      \tilde{v}_{lk}^{(s)}
\label{eq225}\\
&& \quad \quad
+ {E^*}^2 \epsilon_{ilm} D_{mj} \tilde{v}_{lk}^{(s)}
- ({E^*}^2)^\prime \big( \epsilon_{ilm} \frac{k_j k_k}{k} D_{mn} \tilde{v}_{nl}^{(s)}
    - \frac{1}{2} \epsilon_{ijl} \frac{k_m k_n}{k} D_{mn} \tilde{v}_{lk}^{(s)} \big)
    \bigl\} \, \dd^3k \, \dd\omega \, .
\nonumber
\end{eqnarray}
\end{widetext}
Again contributions to $b_{ijk}^{(W)}$ and $b_{ijk}^{(D)}$ with $\delta_{jk}$ have been dropped.

For the calculation of $a_{ij}^{(W)}$, $a_{ij}^{(D)}$, $b_{ijk}^{(W)}$ and $b_{ijk}^{(D)}$
the gradient tensor $\bnab \bmU$ has been considered as a sum of the two parts expressed by $\bW$ and $\bD$.
Of course, such a calculation can also be carried out without splitting $\bnab \bmU$ in this way.
Then a quantity $a^{(\nabla U)}_{ij}$ occurs instead of $a^{(W)}_{ij} + a^{(D)}_{ij}$,
and a quantity $b^{(\nabla U)}_{ijk}$ instead of $b^{(W)}_{ijk} + b^{(D)}_{ijb}$.
We have written equations (\ref{eq219}) and (\ref{eq225}) such that $a^{(D)}_{ij}$ turns into $a^{(\nabla U)}_{ij}$,
and $b^{(D)}_{ijk}$ into $b^{(\nabla U)}_{ijk}$, if on the right--hand sides $D_{lm}$ is replaced by $U_{lm}$.
From these relations for $a^{(\nabla U)}_{ij}$ and $b^{(\nabla U)}_{ijk}$
we can easily derive the relations (\ref{eq217}) and (\ref{eq219})
for $a^{(W)}_{ij}$ and $a^{(D)}_{ij}$ as well as (\ref{eq223}) and (\ref{eq225})
for $b^{(W)}_{ijk}$ and $b^{(D)}_{ijb}$.

\section{Relations for the quantities
${\tilde{\alpha}}_1^{(W)}$, ${\tilde{\alpha}}_1^{(W)}$, $\cdots$ ${\tilde{\kappa}}^{(D)}$}
\label{ftilde}

Analogous to the relations (\ref{eq243}) we have
\begin{widetext}
\begin{eqnarray}
&& {\tilde{\alpha}}_1^{(W)} = (1 / 120) \big(\te^4 \tn^3 (20 \te - \tn) + 4 \te^2 \tn (11 \te^3 + 3 \te^2 \tn
   + 10 \te \tn^2 - 3 \tn^3) \o^2
\nonumber\\
&& \qquad \quad
   + (13 \te^4 + 88 \te^3 \tn - 20 \te^2 \tn^2 + 20 \te \tn^3 + 5 \tn^4) \o^4
   - 4 \te (2 \te - 11 \tn) \o^6 - 5 \o^8 \big) \,
   (\te^2 + \o^2)^{-3} \, (\tn^2 + \o^2)^{-2}
\nonumber\\
&& {\tilde{\alpha}}_2^{(W)} = - (1 / 240) \big( \te^4 \tn^3 (20 \te -13 \tn)
   + 4 \te^2 \tn (3 \te^3 - 11 \te^2 \tn +10 \te \tn^2 + 21 \tn^3) \o^2
\nonumber\\
&& \qquad \quad
   - (31 \te^4 - 24 \te^3 \tn - 140 \te^2 \tn^2 -20 \te \tn^3 +15 \tn^4) \o^4
   + 4 (14 \te^2 + 3 \te \tn -10 \tn^2) \o^6 - 25 \o^8 \big) \, (\te^2 + \o^2)^{-3} \, (\tn^2 + \o^2)^{-2}
\nonumber\\
&& {\tilde{\gamma}}^{(W)} = - (1 / 48) \big( \te^4 \tn^4 + 4 \te^2 \tn (2 \te^3 + 2 \te^2 \tn + 3 \tn^3) \o^2
\nonumber\\
&& \qquad \quad
   + (7 \te^4 + 16 \te^3 \tn + 28 \te^2 \tn^2 - 5 \tn^4) \o^4
   + 4 (4 \te^2 + 2 \te \tn - 3 \tn^2) \o^6 - 7 \o^8 \big) \, (\te^2 + \o^2)^{-3} \, (\tn^2 + \o^2)^{-2}
\nonumber\\
&& {\tilde{\delta}}^{(W)} = (1 / 12) \big(\te^2 - \o^2 \big) \, (\te^2 + \o^2)^{-2}
\nonumber\\
&& {\tilde{\kappa}}^{(W)} = - (1 / {30}) \big(\te^4 \tn^2 - \te^2
(23 \te^2 - 12 \tn^2) \o^2
   -(12 \te^2 + 5 \tn^2) \o^4 - 5 \o^6 \big) (\te^2 + \o^2)^{-3} \, (\tn^2 + \o^2)^{-1}
\nonumber\\
&& {\tilde{\alpha}}^{(D)} = - ( 1 / 120) \big(3 \te^4 \tn^3 (4 \te + 3 \tn)
   - 4 \te^2 \tn (3 \te^3 - 5 \te^2 \tn - 2 \te \tn^2 + 3 \tn^3) \o^2
\label{eq245}\\
&& \qquad \quad
   + (11 \te^4 - 40 \te^3 \tn - 12 \te^2 \tn^2 - 4 \te \tn^3 - 5 \tn^4) \o^4 - 28 \te \tn \o^6 + 5 \o^8 \big)
   (\te^2 + \o^2)^{-3} \, (\tn^2 + \o^2)^{-2}
\nonumber\\
&& {\tilde{\gamma}}^{(D)} = - (1 / 120) \big(3 \te^4 \tn^3 (16 \te - 3 \tn)
   + 4 \te^2 \tn (10 \te^3 + 20 \te \tn^2 + 3 \tn^3) \o^2
\nonumber\\
&& \qquad \quad
   + (9 \te^4 + 64 \te^3 \tn + 52 \te^2 \tn^2 + 32 \te \tn^3 + 5 \tn^4) \o^4
   + 4 (10 \te^2 + 6 \te \tn + 5 \tn^2) + 15 \o^8 \big) \, (\te^2 + \o^2)^{-3} \, (\tn^2 + \o^2)^{-2}
\nonumber\\
&& {\tilde{\beta}}^{(D)} = (1 / 60) \big(\te^4 \tn^3 (10 \te - 3 \tn)
   + 2 \te^2 \tn (\te^3 - 5 \te^2 \tn  + 8 \te \tn^2 - 3 \tn^3) \o^2
\nonumber\\
&& \qquad \quad
   - (7 \te^4 + 16 \te^2 \tn^2 - 6 \te \tn^3 -5 \tn^4) \o^4
   - 2 (5 \te^2 + \te \tn - 5 \tn^2) \o^6 + 5 \o^8 \big) \, (\te^2 + \o^2)^{-3} \, (\tn^2 + \o^2)^{-2}
\nonumber\\
&& {\tilde{\kappa}}^{(D)} = ( 1 / 30) \big(\te^4 \tn^3 (10 \te + 3 \tn)
   + 2 \te^2 \tn (\te^3 + 5 \te^2 \tn  + 8 \te \tn^2 + 3 \tn^3) \o^2
\nonumber\\
&& \qquad \quad
   + (7 \te^4 + 16 \te^2 \tn^2 + 6 \te \tn^3 -5 \tn^4) \o^4
   + 2 (5 \te^2 - \te \tn - 5 \tn^2) \o^6 - 5 \o^8 \big) \, (\te^2 + \o^2)^{-3} \, (\tn^2 + \o^2)^{-2} \, ,
\nonumber
\end{eqnarray}
where $\te$ and $\tn$ stand for $\eta k^2$ and $\nu k^2$,
respectively.
\end{widetext}

{\section{$\bOmega \x \bJ$--effect}} \label{omegaxj}

{For the coefficient $\zeta_1^{(\Omega)}$ defined by (\ref{eq282})
we have according to (\ref{eq243})
\begin{eqnarray}
\zeta_1^{(\Omega)} \!\!\!&=&\!\!\! \frac{64 \pi}{15} \int_{k=0}^\infty \int_{\omega=-\infty}^\infty
     \frac{(\eta k^2)^2 \omega^2}{((\eta k^2)^2 + \omega^2)^2 ((\nu k^2)^2 + \omega^2)}
\nonumber\\
&& \qquad \qquad \qquad \qquad \qquad
    W (k, \omega) \, k^2 \, \dd k \, \dd \omega \, .
\label{eq297}
\end{eqnarray}
Introducing the dimensionless variables $u = (k \lambda_{\mathrm{c}})^2 / q$ and $w = \omega \tau_{\mathrm{c}}$
we find further
\begin{eqnarray}
\zeta_1^{(\Omega)} \!\!\!&=&\!\!\! \frac{32 \pi \tau_{\mathrm{c}}}{15 \lambda^3_{\mathrm{c}}} \, q^{1/2}
    \!\!\int_{u=0}^\infty \int_{w=-\infty}^\infty
    \!\!\frac{u^{5/2} w^2}{(u^2 + \omega^2)^2 (P^2_m u^2 + w^2)}
\nonumber\\
&& \qquad \qquad \qquad
    W ((q u)^{1/2} / \lambda_{\mathrm{c}} , w / \tau_{\mathrm{c}}) \, \dd u \, \dd w
\label{eq299}
\end{eqnarray}
with $P_m = q/p$, and $q$ and $p$ as defined by (\ref{eq261}). We
may assume that $W$ remains finite everywhere. Clearly
$\zeta_1^{(\Omega)}$ always vanishes as $q \to 0$. If $p$ is fixed
the same is obvious for $q \to \infty$. With the reasonable
assumption that $k \, W (k, \omega)$ vanishes as $k \to \infty$ we
can also in the case of fixed $P_m$ conclude that
$\zeta_1^{(\Omega)}$ vanishes as $q \to \infty$.}

\section{$\bW \x \bJ$--dynamo}
\label{wxj}

Consider as in RK03 an infinitely extended fluid with a mean shear flow,
in a Cartesian coordinate system $(x,y,z)$ given by $\bmU = (0, S x, 0)$ with a constant $S$,
and a superimposed turbulence being homogeneous, isotropic, mirror--symmetric and statistically steady
in the limit of vanishing shear.
The only non--zero components of $\bW$ and $\bD$ are then $W_z = S$ and $D_{yx} = D_{yx} = (1/2) S$.
Assume further as in RK03 that $\bmB$ does not depend on $y$.
Then the mean--field induction equation (\ref{eq03}) together with our results for $\bscE$ leads to
\begin{eqnarray}
(\p_t - (\eta + \beta^{(0)}) \Delta ) \mB_x + \delta \, S \, \p^2_{zz} \mB_y &=& 0 \qquad
\nonumber\\
(\p_t - (\eta + \beta^{(0)}) \Delta ) \mB_y - S \, \mB_x - \delta' \, S \, \Delta \mB_x &=& 0 \qquad
\label{eq285}\\
\p_x \mB_x + \p_z \mB_z &=& 0 \qquad
\nonumber
\end{eqnarray}
with
\begin{eqnarray}
\delta &=& \delta^{(W)} - \frac{1}{2} (\kappa^{(W)} - \beta^{(D)} + \kappa^{(D)})
\nonumber\\
\delta' &=& \delta^{(W)} - \frac{1}{2} (\kappa^{(W)} + \beta^{(D)} - \kappa^{(D)}) \, .
\label{eq287}
\end{eqnarray}
The solutions of (\ref{eq11}) are
\begin{equation}
\bmB = \hat{\bmB} \exp (\lambda t + \iu (k_x x + k_z z))
\label{eq289}
\end{equation}
with some constant vector $\hat{\bmB}$ and
\begin{eqnarray}
\lambda &=& - (\eta + \beta^{(0)}) (k^2_x + k^2_z)
\nonumber\\
&& \qquad \qquad
   \pm  |S| \, |k_z| \, \sqrt{\delta \, (1 - \delta' (k^2_x + k^2_z))} \, .
\label{291}
\end{eqnarray}
{We refrain from discussing the case $\delta' (k^2_x + k^2_z) > 1$,
in which the neglect of higher--order derivatives of $\bmB$ in
$\bscE$ could be questionable. Under this restriction} a dynamo can
only exist if $\delta$ is positive.

According to our results (\ref{eq241}) and (\ref{eq245}) for $\delta^{(W)}$, $\kappa^{(W)}$,
$\delta^{(D)}$ and $\kappa^{(D)}$ we have
\begin{eqnarray}
\delta &=& - \frac{\pi}{{15}} \, \int_{k = 0}^\infty \! \int_{\omega
= - \infty}^\infty \!\!
    \Big( \frac{{32} (\eta k^2)^2 \, \omega^2}{((\eta k^2)^2 + \omega^2)^2 ((\nu k^2)^2 + \omega^2)}
\nonumber\\
&& \qquad \qquad \quad
    + \frac{(\eta k^2)^4 + 12 (\eta k^2)^2 \omega^2 - 5 \omega^4}{((\eta k^2)^2 + \omega^2)^3} \Big) \,
\label{eq293}\\
&& \qquad \qquad \qquad \qquad \qquad \qquad \quad
    W (k , \omega) \, k^2 \, \dd k \dd \omega \, .
\nonumber
\end{eqnarray}
Clearly $\delta$ grows monotonically with $\nu$.
Its maximum, $\delta_\mathrm{max}$, is given by
\begin{eqnarray}
\delta_\mathrm{max} &=& - \frac{\pi}{{15}} \,
    \int_{k = 0}^\infty \! \int_{\omega = - \infty}^\infty \!\!\!\!\!
    \frac{(\eta k^2)^4 + 12 (\eta k^2)^2 \omega^2 - 5 \omega^4}{((\eta k^2)^2 + \omega^2)^3} \,
\nonumber\\
&& \qquad \qquad \qquad \qquad \quad
    W (k , \omega) \, k^2 \, \dd k \dd \omega \, .
\label{eq295}
\end{eqnarray}
With an integration by parts with respect to $\omega$ this turns into
\begin{eqnarray}
\delta_\mathrm{max} &=& \frac{\pi}{{15}} \, \int_{k = 0}^\infty \!
\int_{\omega = - \infty}^\infty
    \frac{(\eta k^2)^2 + 5 \omega^2}{((\eta k^2)^2 + \omega^2)^2} \,
    \frac{\p W (k, \omega)}{\p \omega} \,
\nonumber\\
&& \qquad \qquad \qquad \qquad \qquad \qquad
     k^2 \, \omega \, \dd k \dd \omega \, .
\label{eq297}
\end{eqnarray}
It seems reasonable to assume that $\omega \p W / \p \omega \leq 0$.
Then $\delta_\mathrm{max}$ can never be positive. Consequently
$\delta$ is never positive, and a $\bW \x \bJ$ dynamo as considered
above can not work. {This conclusion applies independent of specific
ansatzes like (\ref{eq251}) or (\ref{eq253}).}

\begin{acknowledgments}
We thank N. Kleeorin, I. Rogachevskii and M. Rheinhardt for helpful
comments on the manuscript. {We also thank the helpful anonymous
referees.} R. Stepanov appreciates the financial support from the
grants ISTC--2021 and the BRHE Program.
\end{acknowledgments}

\bibliography{raste}

\end{document}